\documentclass[a4paper, 12pt]{article}
\pdfoutput=1 

\usepackage{jheppub} 

\usepackage[T1]{fontenc} 
\usepackage{blkarray}
\usepackage{relsize}

\usepackage{filecontents}

\usepackage[utf8]{inputenc}
\usepackage{amssymb, amsthm, tabto, epigraph, mathtools, dsfont, verbatim, slashed, mathrsfs}
\usepackage{graphicx, wrapfig}
\usepackage{enumitem}
\usepackage{relsize}
\usepackage{subcaption}
\usepackage{csquotes}
\usepackage{hyperref}
\usepackage{physics}
\usepackage{tabularx}
\usepackage{youngtab}
\usepackage[dvipsnames]{xcolor}
\usepackage{color}
\usepackage[english]{babel}
\usepackage{bbm, dsfont}
\usepackage{multirow}
\usepackage{caption}

\usepackage{float}
\usepackage{bm}

\usepackage{amsmath}
\usepackage{empheq}
 
\makeatletter
\DeclareFontFamily{OMX}{MnSymbolE}{}
\DeclareSymbolFont{MnLargeSymbols}{OMX}{MnSymbolE}{m}{n}
\SetSymbolFont{MnLargeSymbols}{bold}{OMX}{MnSymbolE}{b}{n}
\DeclareFontShape{OMX}{MnSymbolE}{m}{n}{
    <-6>  MnSymbolE5
   <6-7>  MnSymbolE6
   <7-8>  MnSymbolE7
   <8-9>  MnSymbolE8
   <9-10> MnSymbolE9
  <10-12> MnSymbolE10
  <12->   MnSymbolE12
}{}
\DeclareFontShape{OMX}{MnSymbolE}{b}{n}{
    <-6>  MnSymbolE-Bold5
   <6-7>  MnSymbolE-Bold6
   <7-8>  MnSymbolE-Bold7
   <8-9>  MnSymbolE-Bold8
   <9-10> MnSymbolE-Bold9
  <10-12> MnSymbolE-Bold10
  <12->   MnSymbolE-Bold12
}{}

\let\llangle\@undefined
\let\rrangle\@undefined
\DeclareMathDelimiter{\llangle}{\mathopen}%
                     {MnLargeSymbols}{'164}{MnLargeSymbols}{'164}
\DeclareMathDelimiter{\rrangle}{\mathclose}%
                     {MnLargeSymbols}{'171}{MnLargeSymbols}{'171}
\makeatother

\newlength\dlf  

\allowdisplaybreaks

\setlist{nolistsep}
\hypersetup{
    colorlinks=true,
    linkcolor=Mahogany,
    filecolor=Mahogany,      
    urlcolor=Mahogany,
    citecolor=Mahogany
}

\newcommand{\Z}{\mathbb{Z}}

\newcommand{\cD}{\mathcal{D}}

\newcommand{\cH}{\mathcal{H}}

\newcommand{\vol}{\text{Vol}}
\newcommand{\cR}{\mathcal{R}}

\renewcommand{\bar}{\overline}

\theoremstyle{remark}

\theoremstyle{definition}

\usepackage{amssymb}
\usepackage{tikz}
\usetikzlibrary{shapes,arrows,cd,chains,decorations.markings,decorations.pathmorphing,calc}
\tikzset{
->-/.style args={#1rotate#2}{decoration={markings, mark=at position #1 with {\arrow[scale=1.5,rotate = #2 ]{stealth}}}, postaction={decorate}}
}
\tikzset{curve/.style={settings={#1},to path={(\tikztostart)
    .. controls ($(\tikztostart)!\pv{pos}!(\tikztotarget)!\pv{height}!270:(\tikztotarget)$)
    and ($(\tikztostart)!1-\pv{pos}!(\tikztotarget)!\pv{height}!270:(\tikztotarget)$)
    .. (\tikztotarget)\tikztonodes}},
    settings/.code={\tikzset{quiver/.cd,#1}
        \def\pv##1{\pgfkeysvalueof{/tikz/quiver/##1}}},
    quiver/.cd,pos/.initial=0.35,height/.initial=0}
\tikzset{tail reversed/.code={\pgfsetarrowsstart{tikzcd to}}}
\tikzset{2tail/.code={\pgfsetarrowsstart{Implies[reversed]}}}
\tikzset{2tail reversed/.code={\pgfsetarrowsstart{Implies}}}

\newmuskip\pFqskip
\mathchardef\pFcomma=\mathcode`, 

\graphicspath{{./images/}}

\renewcommand{\bar}{\overline}

\usepackage{braket}

\usepackage[export]{adjustbox}

\title{\center{Monopole-Fermion Scattering from Twisted Conformal Boundary States}}

\abstract{We derive monopole-fermion scattering correlators from conformal boundary states in twisted sectors. The key ingredient is that symmetry-preserving boundary conditions contain sectors twisted by topological symmetry lines. Their overlaps determine correlators of operators attached to these lines, including the outgoing excitations required by charge conservation in the Callan-Rubakov effect. We construct the twisted boundary states of $N$ free Dirac fermions with $U(1)^N$-preserving boundary conditions, extending the familiar description in terms of charge lattices to shifted lattices describing twisted sectors. Using cluster decomposition and open-closed duality, we determine their coefficients and fix their normalization relative to the untwisted boundary state. We find that the Callan-Rubakov boundary condition yields a two-point function between an incoming fermion and an outgoing defect operator with a coefficient of unit modulus, even when the corresponding correlators of ordinary local fermions vanish. Finally, we relate these correlators to four-dimensional partial-wave scattering using gauge-covariant asymptotic states. We identify the missing $s$-wave contribution to the unitarity relation and argue that the twisted channel supplies it with precisely the required normalization.}

\author[1]{Clay C\'ordova,}
\author[1,2]{Diego Garc\'ia-Sep\'ulveda,}
\author[1]{Lian-Tao Wang,}
\author[3]{and Yen-Chieh Wang}

\affiliation[1]{Leinweber Institute for Theoretical Physics \& Enrico Fermi Institute, University of Chicago, Chicago, IL 60637, USA}

\affiliation[2]{Society of Fellows, Harvard University, Cambridge, MA 02138, USA}

\affiliation[3]{Mani L. Bhaumik Institute for Theoretical Physics, University of California Los Angeles, Los Angeles, CA 90095, USA}

\emailAdd{clayc@uchicago.edu}
\emailAdd{dgarciasepulveda@fas.harvard.edu}
\emailAdd{liantaow@uchicago.edu}
\emailAdd{yenchieh@ucla.edu}

\begin{document}

\maketitle

\section{Introduction}

Magnetic monopoles are among the most robust nonperturbative predictions of grand unification: when a unified gauge group is spontaneously broken to a subgroup containing electromagnetism, topologically stable 't~Hooft-Polyakov monopoles generically arise \cite{tHooft:1974Monopole,Polyakov:1974Monopole, GoddardOlive:1978,Preskill:1984Monopoles}. For symmetry-breaking scales
characteristic of grand unified theories (GUTs), these monopoles are typically superheavy. Depending on their production mechanism and the evolution history of the Universe, some of them may survive as relics from the early Universe. This offers the possibility of probing their interactions~\cite{Preskill:1979Monopoles, Preskill:1984Monopoles}. The Callan-Rubakov effect provides a striking example of such a probe~\cite{Rubakov:1981Monopole,Rubakov:1982fp,Callan:1982ah, Callan:1982au,Callan:1983Catalysis}. Scattering of light fermions from a GUT monopole can induce baryon- and lepton-number violating processes, in particular monopole-catalyzed nucleon decay. The amplitudes of such processes are not suppressed by inverse powers of the unification scale \cite{Rubakov:1981Monopole,Rubakov:1982fp,Callan:1982ah, Callan:1982au,Callan:1983Catalysis,RubakovSerebryakov:1983, Rubakov:1988aq}.

The mechanism has been studied explicitly in unified theories such as $SU(5)$ and $SO(10)$~\cite{RubakovSerebryakov:1983,DawsonSchellekens:1983}, and has a wide range of phenomenological consequences. Monopoles captured in astrophysical bodies can catalyze repeated nucleon decays, leading to constraints from stellar energetics and to characteristic neutrino
signals~\cite{DimopoulosPreskillWilczek:1982,KolbColgateHarvey:1982, SuperK:2012Monopole}. Similarly, monopoles traversing terrestrial detectors can produce a sequence of nucleon-decay events along their trajectories. These signatures have motivated dedicated searches in Soudan, IMB, MACRO, Super-Kamiokande, and IceCube~\cite{Soudan:1987Monopole,IMB:1994Monopole, MACRO:2002Catalysis,SuperK:2012Monopole,IceCube:2014Monopole}, and have recently been reconsidered for future experiments including DUNE and Hyper-Kamiokande~\cite{Candela:2025Monopoles}. The same unsuppressed baryon-number violation also suggests a possible role for the Callan-Rubakov process in baryogenesis. The potential connection between monopoles and the cosmological baryon asymmetry was recognized soon after the discovery of monopole-catalyzed baryon-number violation~\cite{Nussinov:1982BaryonAsymmetry,Ellis:1982CatalysisBAU}. The large Callan-Rubakov rate can have two complementary consequences: monopole-fermion scattering can wash out a previously generated baryon asymmetry, thereby placing cosmological constraints on the monopole abundance and catalysis rate~\cite{Ellis:1982CatalysisBAU}, while, in the presence of suitable $CP$ violation and departure from equilibrium, the same baryon-number-violating process can itself participate in baryogenesis. Early
scenarios along these lines were explored in the context of monopole annihilation and, in particular, the Langacker-Pi mechanism~\cite{LangackerPi:1980,Davis:1992MonopoleBaryogenesis}; more recently, it has been shown explicitly that a $CP$-violating $\theta$ angle can bias Callan-Rubakov scattering in minimal $SU(5)$ and generate the observed baryon asymmetry~\cite{Brennan:2024MonopoleBaryogenesis}.

In all of these potential implications for particle physics, the effect depends directly on the monopole-fermion scattering probabilities, so a first-principles determination of the Callan-Rubakov scattering amplitudes and cross sections is important for connecting the underlying field theory to experimental searches.

In this paper we study monopole-fermion scattering using twisted conformal boundary states. For a minimally charged monopole, the lowest angular momentum modes propagate radially without a centrifugal barrier. In the massless, weak-coupling approximation, their dynamics reduce to free fermions on a half-line, with the monopole core represented by a boundary condition at the origin~\cite{Callan:1982au,Callan:1983ed,Polchinski:1984uw}. This description underlies the fermion-rotor model~\cite{Polchinski:1984uw} and the boundary conformal field theory (BCFT) treatment of monopole catalysis developed in~\cite{Affleck:1993np}.

The boundary condition is nevertheless subtle: conservation of the preserved currents can require outgoing excitations with charges that no ordinary local fermion operator can carry. Restricting to the original Fock space then produces an apparent failure of unitarity. For four fermion flavors, the missing states were identified with collective excitations transforming in a spinor representation of $SO(8)$~\cite{Maldacena:1995pq}. More recent work interprets the outgoing states through a varying Fock space~\cite{Hamada:2022eiv} and operators attached to topological symmetry defects~\cite{vanBeest:2023dbu,vanBeest:2023mbs}. In two dimensions, such an operator lies at the end of a topological line and belongs to its twisted sector. Our goal is to directly use this formalism to extract the relevant
correlation functions, apply the LSZ reduction formula to determine a scattering matrix, and explicitly verify unitarity.

We first recall the ordinary boundary state construction~\cite{Ishibashi:1988kg,Cardy:1989ir,CARDY1991274,recknagel_schomerus_2013}.
Quantizing a theory on the interval with specified boundary conditions gives a Hilbert space of states, whose trace is the cylinder partition function. By contrast, viewing the cylinder as a circle propagated through time represents the same partition function as an overlap of states in the bulk Hilbert space on a circle. The conformal gluing conditions admit solutions, called Ishibashi states, which pair compatible left- and right-moving conformal families. A physical boundary state is an appropriate linear combination of these solutions. Its coefficients determine bulk one-point functions, and are constrained by consistency between the two quantizations, known as open-closed duality.

A much less familiar aspect of this description is that a symmetric boundary also defines boundary states in twisted sectors. Suppose a topological symmetry line $L$ can end on the boundary $B$ at a topological junction. In the closed channel, the line intersects the spatial circle, so this configuration prepares a state $|B_L\rangle$ in the twisted Hilbert space $\mathcal H_L$. Thus, the same boundary condition naturally defines a family of states in different sectors once the corresponding line endpoints are specified, with $|B_1\rangle=|B\rangle$~\cite{Fukusumi:2021zme,Ebisu:2021acm,Choi:2023xjw,Choi:2024tri}. These states encode the boundary's couplings to defect operators. They are needed already in the original, ungauged theory whenever such operators appear in a correlator.

The ingredients of this construction have a long history in the study of twisted partition functions and disorder operators~\cite{Petkova:2000ip}. Twisted boundary states arise naturally in orbifold constructions and fermionization, where they relate boundary conditions across dualities~\cite{Fukusumi:2021zme,Ebisu:2021acm,Watts:2000kp}. More recently, they have been used to resolve partition functions and entanglement according to symmetry charge~\cite{Kusuki:2023bsp,Choi:2024wfm}. A general framework incorporating non-invertible symmetries and their boundary junctions was developed in~\cite{Choi:2024tri}. Here we emphasize the role of twisted boundary states in determining scattering amplitudes, for which the relative normalization of twisted and untwisted contributions is essential.

Our approach complements recent studies of outgoing wavepackets~\cite{Tachikawa:2026cxd}, scattering through non-invertible defects~\cite{Ueda:2025ecm,Wei:2026fsn,Arias-Tamargo:2026urw}, and defect anomalies~\cite{Antinucci:2026uuh}, as well as extensions to non-abelian rotors and chiral symmetries~\cite{Koundouros:2026rnb,Ohmori:2026tip,Arias-Tamargo:2026kbv}. See also \cite{Kitano:2021pwt,Loladze:2024ayk, Csaki:2020inw, Csaki:2021ozp, Csaki:2024ajo}.

Concretely, an annulus with a symmetry line joining its boundaries admits the two descriptions
\begin{equation}
    \operatorname{Tr}_{\mathcal H_{AB}}
    \!\left(Lq^{L_0-c/24}\right)
    =
    \langle A_L|
    \widetilde q^{\frac12(L_0+\bar L_0-c/12)}
    |B_L\rangle.
\end{equation}
Here $\mathcal H_{AB}$ is the Hilbert space on an interval with boundaries $A$ and $B$, and $q$ and $\widetilde q$ are the open- and closed-channel parameters. In the open channel, the trace weights boundary operators by their symmetry eigenvalues. In the closed channel, the boundary states are expanded in Ishibashi states belonging to $\mathcal H_L$. Matching these expressions gives the twisted analogue of the Cardy condition~\cite{Fukusumi:2021zme,Kusuki:2023bsp,Choi:2024tri}. Unlike ordinary state multiplicities, the coefficients in the symmetry-weighted trace can have signs or phases; these record the symmetry action and constrain the twisted boundary couplings.

The corresponding overlaps determine correlators of defect endpoint operators. For a normalized scalar primary $\tilde\varphi_L$ of weights $(h,h)$, with its line joining the insertion to the specified endpoint on $B$, one finds
\begin{equation}
    \bigl\langle\tilde\varphi_L(z,\bar z)\bigr\rangle_B
    =
    \frac{\langle B_L|\tilde\varphi_L\rangle}
         {\langle B|0\rangle}
    \frac{1}{|z-\bar z|^{2h}}.
\end{equation}
The numerator probes the twisted sector, while the denominator remains the untwisted vacuum overlap that normalizes the path integral. An overlap with the ordinary boundary state cannot recover this amplitude, since the inserted state belongs to a different sector. Using the appropriate twisted boundary state therefore extends the familiar boundary state method to precisely the operators required in the scattering problem.

A simple example is the free boundary of the critical Ising model. Its ordinary boundary state contains the identity and energy conformal families. Nevertheless, the spin-flip line can end on this boundary, and its twisted boundary state couples to the disorder field. The resulting nonzero disorder one-point function, familiar from Kramers--Wannier duality~\cite{Watts:2000kp}, follows from the twisted overlap and its normalization. We review this example in Appendix \ref{IsingExample}. It illustrates how a boundary can have non-trivial couplings that are invisible in its untwisted boundary state.

For the monopole application, we begin with the $U(1)^N$-preserving boundary states of $N$ Dirac fermions~\cite{Smith:2019jnh}. They are characterized by a rational orthogonal current-gluing matrix $\cR$ and the charge lattice $\Lambda[\cR]=\{\lambda\in\mathbb Z^N:\cR\lambda\in\mathbb Z^N\}$. The corresponding twisted states are constructed from shifted lattices $\Lambda[\cR]+a$, where $a\in\mathbb Z^N$ labels a finite abelian symmetry defect. Current gluing, cluster decomposition, and twisted open--closed duality constrain their coefficients and fix their normalization relative to the untwisted state. For the Callan--Rubakov boundary, $\cR_{ij}=\delta_{ij}-2/N$, and an incoming fermion of charge $e_i$ must be paired in the boundary two-point function with an antiholomorphic defect operator of charge $-\cR e_i$. The resulting two-point function has a coefficient of unit modulus. This gives a direct BCFT derivation of the dressed-propagator normalization found in the fermion--rotor description~\cite{Loladze:2024ayk,Loladze:2025jsq}.

Finally, we connect the twisted correlators to four-dimensional scattering. Using gauge-covariant asymptotic states, we isolate the missing $s$-wave contribution to the partial-wave unitarity relation. We derive how it relates to the BCFT correlator for boundary conditions acting linearly on the fermions, and argue that the same identification for the twisted channel supplies exactly the missing contribution. In this way, boundary state consistency gives a quantitative account of probability conservation.

In Section \ref{Section2} we review boundary states in the presence of symmetry defects. Section \ref{Section3} constructs the twisted states and correlators for free Dirac fermions, and Section \ref{Section4} develops their relation to four-dimensional scattering. The appendices collect the partial-wave analysis and the Ising example.

\section{Twisted Boundary Conformal Field Theory} \label{Section2}

In this section we describe the formalism of boundary CFT and global symmetries. In particular, we discuss the contributions of twisted sectors of the Hilbert space to boundary states, which are central to our application to monopole physics. See e.g.\ \cite{DiFrancesco:1997nk} for an overview of background material on 2$d$ CFTs.

\subsection{Topological Defect Lines} \label{TDLs}

Global symmetries in 2$d$ CFTs may be viewed as topological defect line operators in the theory \cite{Gaiotto:2014kfa, Bhardwaj:2017xup, Chang:2018iay, Lin:2019kpn, Lin:2021udi}. In this work we will only need to consider the case where the symmetry is a finite abelian group e.g.\ $\mathbb{Z}_{N}$, and as such restrict the discussion to that case. Generalizations to non-invertible symmetries appear in \cite{Frohlich:2004ef, Lin:2023uvm, Thorngren:2021yso}.

\begin{figure}[t]
\centering
        \includegraphics[scale=1.4]{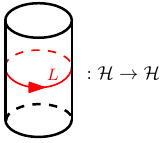} 
        \caption{A topological line defect placed at a fixed time around the spatial direction acts as an operator on the Hilbert space of the theory quantized on $S^{1}.$ By the operator-state correspondence, this is the space of local operators.} \label{TopologicalLineAction}
\end{figure}

The most familiar aspect of a global symmetry is that it acts on the space of local operators. For instance, if the topological line operator generates a $\mathbb{Z}_{N}$ symmetry, we can organize the operators according to their $\mathbb{Z}_{N}$ charge. By the operator-state correspondence, this is equivalent to studying the action of the symmetry on the Hilbert space of states quantized on the circle, $\mathcal{H},$ as shown in Figure \ref{TopologicalLineAction}.

\begin{figure}[!b]
\centering
        \includegraphics[scale=1.4]{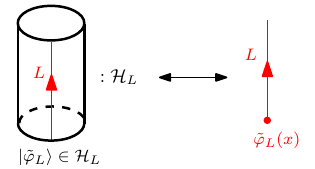} 
        \caption{Quantizing a 2$d$ CFT on a circle with a topological line defect $L$ extended in the time direction gives rise to a defect Hilbert space $\mathcal{H}_{L}$. Under the operator-state correspondence, states in the defect Hilbert space $|\tilde{\varphi}_{L} \rangle \in \mathcal{H}_{L}$ are mapped to quasi-local operators $\tilde{\varphi}_{L}(x)$ at the end of the topological defect line $L$.} \label{Cylinder-TwistedSector}
\end{figure}

Viewed in spacetime, the description above corresponds to inserting the topological defect line on a spatial slice at fixed time.  Instead, we can also consider the defect extended in time, intersecting the spatial circle at a point. This arrangement modifies the boundary condition around the circle and defines a ``twisted'' Hilbert space $\mathcal{H}_{L}$. A state in this twisted Hilbert space $|\tilde{\varphi}_{L} \rangle \in \mathcal{H}_{L}$ is mapped through the operator-state correspondence to a quasi-local operator characterized by the fact that it is located at the end of the topological line $L$, as depicted in Figure \ref{Cylinder-TwistedSector}.

The spectrum of the twisted Hilbert space $\mathcal{H}_{L}$ is closely connected to that of the untwisted sector $\mathcal{H}$ by modularity. One may define a partition function that counts the number of states in the twisted Hilbert space $\mathcal{H}_{L}$, weighted by an exponential of their energy and spin:
\begin{equation}
    \mathcal{Z}_{L}(\tau,\bar{\tau}) = \mathrm{Tr}_{\mathcal{H}_{L}}[q^{L_{0} - c/24} \bar{q}^{\bar{L}_{0} - \bar{c}/24}],
\end{equation}
where $q = e^{2 \pi i \tau}$, and $c$ is the central charge. Geometrically, this partition function corresponds to a correlator on the torus with no insertions of local operators, but with the topological defect line $L$ wrapping a cycle that we view as the time direction. By modularity, one can exchange the roles of the cycles of the torus, with the compact Euclidean time direction now running perpendicular to $L$. Then, one finds that:
\begin{equation}
    \mathcal{Z}_{L}\left( -\frac{1}{\tau},-\frac{1}{\bar{\tau}} \right) = \mathrm{Tr}_{\mathcal{H}}[L \, q^{L_{0} - c/24} \bar{q}^{\bar{L}_{0} - \bar{c}/24}].
\end{equation}
Thus, via modular invariance, the spectrum of states in the twisted sector $\mathcal{H}_{L}$ is determined by the action of the line $L$ on the states of the untwisted sector $\mathcal{H}$.

\subsection{Untwisted Boundary Conformal Field Theory}

  Our next ingredient is boundary conformal field theory.  Specifically, this means that we describe a 2$d$ CFT in the presence of a boundary condition that preserves conformal symmetry. First, we summarize the theory of conformal boundary conditions in the absence of topological defect lines using standard formalism \cite{Cardy:1989ir, CARDY1991274, DiFrancesco:1997nk, recknagel_schomerus_2013}.

\begin{figure}[t]
\centering
        \includegraphics[scale=1.25]{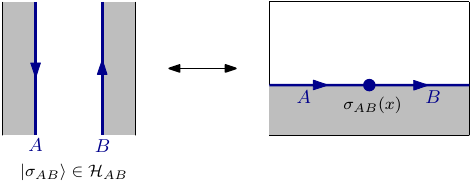} 
        \caption{The operator-state correspondence maps every state $| \sigma_{AB} \rangle$ in the Hilbert space $\mathcal{H}_{AB}$ of the 2$d$ CFT defined on the interval with conformal boundary conditions $A$ and $B$ to a local operator $\sigma_{AB}(x)$ located at the boundary of the half-plane, creating a junction between the boundary conditions $A$ and $B$. Such operators are also called \emph{boundary-changing operators.}} \label{BoundaryOperatorStateCorrespondence}
\end{figure}

We place our 2$d$ CFT on a strip, with possibly different conformal boundary conditions on each end. The boundary conditions are denoted by capital letters $A$ and $B$. See, for example, the configuration on the left in Figure \ref{BoundaryOperatorStateCorrespondence}.  Given such boundary conditions the 2$d$ CFT defines a new Hilbert space of states, $\mathcal{H}_{AB}$, on the interval.  By conformal invariance we can map the strip to the upper half-plane, and the boundary of the strip is mapped to the real line bounding the upper half-plane. The operator-state correspondence then relates each state $| \sigma_{AB} \rangle \in \mathcal{H}_{AB}$ to a local boundary operator $\sigma_{AB}(x)$ that separates the conformal boundary conditions $A$ and $B$ on the real line.  In this correspondence the energy of the state $| \sigma_{AB} \rangle \in \mathcal{H}_{AB}$ is mapped to the scaling dimension of the operator $\sigma_{AB}(x)$.  See Figure \ref{BoundaryOperatorStateCorrespondence}.

The complete spectrum of the theory on the strip is concisely encapsulated in the partition function:
\begin{equation}
    \mathcal{Z}_{AB}(\beta) = \mathrm{Tr}_{\mathcal{H}_{AB}}[\exp(- \beta H)].
\end{equation}
Alternatively, via the correspondence above, this partition function can be viewed as a sum over boundary-changing operators weighted by their scaling dimension, which is the eigenvalue of the Virasoro generator $L_{0}.$ Then the partition function reads:
\begin{equation}
    \mathcal{Z}_{AB}(q) = \mathrm{Tr}_{\mathcal{H}_{AB}}[q^{L_{0} - c/24}],
\end{equation}
where $q = e^{-\beta}$.

\begin{figure}[t]
\centering
        \includegraphics[scale=1.6]{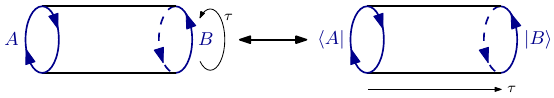} 
        \caption{Modular invariance relates the partition function of the theory on a cylinder with boundary conditions $A$ and $B$ to a corresponding transition amplitude between boundary states. } \label{CylinderPartitionFunction}
\end{figure}

Geometrically, the partition functions above may be viewed as a path integral on a cylinder where we have chosen a slicing such that time is periodic and space is an interval.  Invoking modular invariance in this geometry then provides a rather practical way to characterize a large class of boundary conditions. Swapping the roles of space and time we find that the partition function must also have a description in terms of a transition amplitude between states $| A \rangle$ and $|B \rangle$ describing, respectively, the boundary conditions $A$ and $B$.  See Figure \ref{CylinderPartitionFunction}.  The states $| A \rangle$ and $|B \rangle$ are called boundary states and are constructed from states of the 2$d$ CFT quantized on a circle, denoted $\mathcal{H}$ above.\footnote{Boundary states are formal, non-normalizable states.}  Expressed in this manner we have:
\begin{equation} \label{CardyCondition}
    \mathcal{Z}_{AB}(q) = \langle A | \tilde{q}^{\frac{1}{2}(L_{0} + \bar{L}_{0} - c/12)} | B \rangle,
\end{equation}
where $\tilde{q} = e^{-4\pi^{2}/\beta}$. This setup based on boundary states to describe conformal boundary conditions is known as the ``closed-channel'' description, while the original setup based on states on the strip is called the ``open-channel'' description.

The description in terms of boundary states gives a transparent way to impose symmetries on the boundary conditions: we simply impose them on the boundary states.  For instance, we have throughout been assuming conformal invariance of the boundary conditions, which means that the following combination of Virasoro modes annihilates the boundary states:
\begin{equation} \label{gluingcondition}
    (L_{n} - \bar{L}_{-n})| A \rangle = 0, \quad \forall n \in \mathbb{Z},
\end{equation}
where $L_{n}$ ($\bar{L}_{n}$) is the holomorphic (antiholomorphic) mode of the stress-energy tensor for the theory quantized on a circle, and an identical condition holds for $|B\rangle.$

Similarly, a given conformal boundary condition may also preserve some of the internal symmetries of the theory. For instance, if the CFT has a continuous symmetry encoded by some spin-one conserved currents $J(z)$, $\overline{J}(\bar{z})$, we can ask for a diagonal combination of these currents to be preserved by the boundary condition. This results in a corresponding gluing condition for the boundary state:
\begin{equation} \label{internalgluingcondition}
    (J_{n} + \Omega(\overline{J}_{-n})) | A \rangle = 0,
\end{equation}
where $J_{n}$ ($\overline{J}_{n}$) are the modes for the above currents, and $\Omega$ is an automorphism of the symmetry algebra dictating which diagonal combination of currents is preserved. Of course, the most general conformal boundary condition need not preserve any internal symmetry. However, boundary conditions preserving enhanced symmetry are often the simplest to characterize fully.

The gluing conditions \eqref{gluingcondition} and \eqref{internalgluingcondition} are linear constraints on the boundary states and are straightforward to solve making use of so-called \textit{Ishibashi states} \cite{Ishibashi:1988kg, Cardy:1989ir}. Such states are denoted as $| i \rrangle$, and are expressed as an infinite summation of states consisting of a (chiral) primary $i$ and its infinite number of (chiral) descendants:
\begin{equation}
    | i \rrangle = \sum_{N = 0}^{\infty} |i; N \rangle \otimes U \overline{|i; N \rangle},
\end{equation}
where $|i; N \rangle$, $N \in \mathbb{Z}^{\geq 0}$ denotes an orthonormal basis of the chiral algebra in the sector corresponding to the (chiral) primary $i$ and its descendants.\footnote{This expression holds for trivial automorphism $\Omega$. For non-trivial $\Omega$ the corresponding Ishibashi state differs by the action of $\Omega$ over the antiholomorphic sector. For the precise expression see \cite{recknagel_schomerus_2013}, although this will not be too important for our purposes.} Ishibashi states are thus labeled by the (chiral) primary labels $i$.

Importantly, Ishibashi states are characterized by the fact that
\begin{equation}
    \llangle i | q^{L_{0} - c/24} | j \rrangle = \delta_{i,j} \chi_{i}(q),
\end{equation}
where $\chi_{i}(q)$ corresponds to the character of the primary $i$ with respect to the currents preserved by the boundary. See \eqref{internalgluingcondition}.

Ishibashi states themselves are not boundary states, but they serve as a tool from which a boundary state can be constructed as a linear combination of Ishibashi states:
\begin{equation} \label{UntwistedBoundaryState}
    | A \rangle = \sum_{i} \mathcal{C}_{A}^{\ i} | i \rrangle.
\end{equation}
The gluing conditions \eqref{gluingcondition}, \eqref{internalgluingcondition} are then automatically satisfied, and the main complication is finding the coefficients $\mathcal{C}_{A}^{i}$. One important constraint arises from Eqn. \eqref{CardyCondition}. Evaluating the overlap, we find that 
\begin{equation} \label{CardyConditionExplicit}
    \sum_{i} \mathcal{C}^{\ i*}_{A} \, \mathcal{C}^{\ i}_{B} \, \chi_{i}(\tilde{q}) = \sum_{i} n_{AB}^{i} \chi_{i}(q), \quad n^{i}_{AB} \in \mathbb{Z}, \quad n^{i}_{AB} \geq 0.
\end{equation}
This is, following a modular transformation, the overlap should be suitably interpreted as a partition function for the theory defined on a strip. Consequently, the coefficients $\mathcal{C}^{i}_{A}$ must be chosen so that the partition function may be expressed as an expansion involving coefficients $n^{i}_{AB}$, which must be nonnegative integers. This requirement that the overlap possess a suitable partition function interpretation is known as the \textit{Cardy condition} and provides a strong constraint on the form that boundary states can take.

Let us next discuss the correlators of bulk local operators in the presence of a conformal boundary condition. We provide here a quick discussion that is sufficient for our purposes. For more details, see \cite{recknagel_schomerus_2013}. 

A correlator of local bulk operators in the presence of a boundary $A$ (open channel) can be calculated from a conformal transformation that relates the unit circle (closed channel) and the lower half-plane with the boundary on the real line. The explicit conformal transformation reads:
\begin{equation} \label{FromUnitCircletoLowerHalfPlane}
    \zeta = \frac{1-iz}{1+iz}, \quad \bar{\zeta} = \frac{1 + i \bar{z}}{1 - i \bar{z}},
\end{equation}
where $\zeta$ stands for the coordinates in the unit circle (closed channel) and $z$ stands for the coordinates in the (lower) half-plane.

In the unit circle description, the boundary condition is represented by a boundary state at the edge of the circle. Here we can calculate correlators as in standard 2$d$ CFT on the sphere: taking the bulk OPE of the bulk local operators and computing the overlap with the final state, which in our case is a boundary state. The above conformal transformation then allows us to recover the corresponding correlator in the open channel, with the theory defined on the lower half-plane and the boundary condition on the real line. This is
\begin{equation} \label{CorrelatorsandBoundarystates}
    \langle A | \varphi_{1}(\zeta_{1}, \bar{\zeta}_{1}) \cdots \varphi_{N}(\zeta_{N}, \bar{\zeta}_{N} ) | 0 \rangle = J(\zeta,z) \langle \varphi_{1}(z_{1}, \bar{z}_{1}) \cdots \varphi_{N}(z_{N}, \bar{z}_{N} ) \rangle_{A},
\end{equation}
where $J(\zeta, z)$ is the Jacobian from the transformation in between the unit circle ($\zeta$ coordinates) and the half-plane ($z$ coordinates).

We can apply this now to calculate e.g. one-point functions of (scalar) primaries. Consider a scalar primary $\varphi_{i}$ on the unit circle ``surrounded by'' a boundary state $| A \rangle$. The corresponding overlap $\langle A | \varphi_{i}(\zeta, \bar{\zeta}) | 0 \rangle$ can be calculated from the Ward identities implied by conformal symmetry. The result is:
\begin{equation} \label{onepointcircleoverlap}
    \langle A | \varphi_{i}(\zeta, \bar{\zeta}) | 0 \rangle =   \frac{ \mathcal{C}_{A}^{\ i} }{(1-\zeta \bar{\zeta})^{2h_{i}}},
\end{equation}
where the normalization here is fixed by the standard convention for bulk two-point functions, $\langle i| \varphi_{i}(\zeta, \bar{\zeta})|0\rangle =1$, with $|i\rangle$ the state corresponding to the $\varphi_{i}$ primary. Thus, only the boundary state contributes a non-trivial factor.

\begin{figure}[t]
\centering
        \includegraphics[scale=1.7]{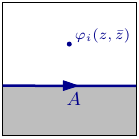} 
        \caption{One-point function of a local operator $\varphi_{i}(z,\bar{z})$ in the presence of a conformal boundary condition $A$.} \label{BoundaryOnePoint}
\end{figure}

Using \eqref{CorrelatorsandBoundarystates}, we find the one-point function in the presence of a boundary condition $A$ on the half-plane:
\begin{equation}  \label{untwistedonepointfunction}
    \langle \varphi_{i}(z, \bar{z}) \rangle_{A} =  \frac{\langle A | i \rangle / \langle A | 0 \rangle}{|z - \bar{z}|^{2h_{i}}},
\end{equation}
where we have chosen a normalization such that the local operator corresponding to the identity operator always has a unit one-point function. A depiction of this configuration is shown in Figure \ref{BoundaryOnePoint}. Notice that the $z$-dependence just obtained could have been determined via conformal invariance on the half-plane, which would have fixed the one-point function up to a constant. The prior argument fixes the constant to be the corresponding coefficient in the boundary state. See also \cite{CARDY1991274}.

To compute higher-point correlators of bulk local operators in the presence of a boundary, one first computes the corresponding overlap in the unit circle (closed channel) by computing the bulk OPE of local operators until one reduces everything to overlaps involving a single field. This can then be evaluated via \eqref{onepointcircleoverlap}. Once this is done, \eqref{CorrelatorsandBoundarystates} determines the corresponding correlator on the half-plane. 

An important constraint over correlators is a cluster decomposition constraint (see \cite{Smith:2019jnh}). This says that, as we approach the unit circle:
\begin{equation} \label{ClusterDecomposition}
    \lim_{|z|, | \omega| \to 1}\frac{\langle\varphi_{1}(z)\varphi_{2}(\omega)\rangle_{A}}{\langle \varphi_{1}(z)\rangle_{A}\langle\varphi_{2}(\omega)\rangle_{A}} = 1.
\end{equation}
Essentially, what this constraint states is that when the local operators approach the boundary of the half-plane, we can use the conformal transformations to take local operators very far apart and the two-point correlator must decompose as standard under cluster decomposition.

\begin{figure}[!b]
\centering
        \includegraphics[scale=1.7]{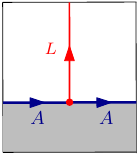} 
        \caption{A conformal boundary condition (blue) symmetric under a topological defect line $L$ (red) allows for $L$ to end at a topological junction at the boundary.} \label{TwistedConformalBoundary}
\end{figure}

\subsection{Twisted Boundary Conformal Field Theory} \label{TwistedBCFT}

Thus far we have discussed the theory of conformal boundary conditions in the absence of any topological defect lines. We now address the theory of conformal boundary conditions ``in a twisted sector,'' referring to the theory of conformal boundary conditions in the presence of topological defect lines ending at a topological junction on the boundary. For recent work on BCFT on twisted sectors, see \cite{Fukusumi:2021zme, Ebisu:2021acm, Collier:2021ngi, Choi:2023xjw, Choi:2024tri, Choi:2024wfm}. See Figure \ref{TwistedConformalBoundary} for a depiction of the configuration of interest.

\begin{figure}[t]
\centering
        \includegraphics[scale=1.5]{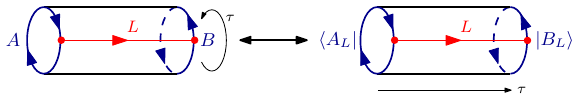} 
        \caption{Modular invariance relates the partition function dressed by the action of the topological defect line $L$ to a corresponding overlap of boundary states on the twisted sector of the closed channel.} \label{TwistedCylinderPartition}
\end{figure}

As before, modular invariance provides a concrete way of studying boundary conditions. This time, we are interested in a configuration with a topological line defect extending from the left boundary of the strip to the right boundary. By modularity, we can again reinterpret the configuration as an overlap of states quantized on a circle, but this time with the topological defect line extended through time puncturing the circle. See Figure \ref{TwistedCylinderPartition} for a pictorial representation of this equivalence. This is, we must have the following equality:
\begin{equation} \label{twistedmodularity}
    \mathrm{Tr}_{\mathcal{H}_{AB}}[L \, q^{L_{0} - c/24}] =  \langle A_{L} | \tilde{q}^{\frac{1}{2}(L_{0} + \bar{L}_{0} - c/12)} | B_{L} \rangle,
\end{equation}
where on the left-hand side the topological line $L$ acts over the states of the theory in the strip with boundary conditions $A$ and $B$. On the right-hand side, we reinterpret the configuration as an overlap of ``twisted'' boundary states $| A_{L} \rangle$ and $| B_{L} \rangle$. Notice that since the topological defect line is now extended in time, these twisted boundary states are constructed from states in the twisted sector $\mathcal{H}_{L}$ defined above in Section \ref{TDLs}. Compare this setup with the one in the previous section. Clearly, when $L$ is the trivial line defect we recover the standard boundary state formalism.

The next steps are now analogous to those in the untwisted case. First, to solve the gluing conditions, we define Ishibashi states in the twisted sector. Namely, states that are defined as a sum over a primary and all the (chiral) descendants, but this time with respect to the states in the twisted sector:
\begin{equation}
    |i_{L} \rrangle \in \mathcal{H}_{L},
\end{equation}
where $\mathcal{H}_{L}$ stands for the Hilbert space in the sector twisted by $L$. Of course, standard Ishibashi states are obtained by setting $L$ to be the identity line.

As before, actual twisted boundary states $|A_{L} \rangle$ are constructed as linear combinations of Ishibashi states. We have:
\begin{equation} \label{TwistedBoundaryState}
    | A_{L} \rangle = \sum_{i} \mathcal{C}(L)_{A}^{\ i} | i_{L} \rrangle.
\end{equation}
Once we have this expansion, Eqn. \eqref{twistedmodularity} provides strong constraints over the coefficients $\mathcal{C}(L)_{A}^{\ i}$. Specifically, the $\mathcal{C}(L)_{A}^{\ i}$ must be selected so that, after a modular transformation, we obtain a partition function that matches the one on the right-hand side of \eqref{CardyConditionExplicit}, but with the action of $L$ applied to each state.

\begin{figure}[t] \hspace{2.4cm}
\begin{subfigure}{.45\textwidth}
  \includegraphics[scale=1.4]{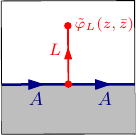}
\end{subfigure}%
\hspace{-1.0cm}
\begin{subfigure}{.45\textwidth}
  \includegraphics[scale=1.4]{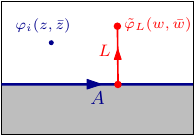}
\end{subfigure}
\caption{On the left: One-point correlator of a quasi-local operator $\tilde{\varphi}_{L}(z,\bar{z})$ living at the end of a topological defect line $L$ in the presence of a conformal boundary condition $A$ that is symmetric under $L$. On the right: Two-point correlator of a genuine local operator $\varphi_{i}(z,\bar{z})$ and the quasi-local operator $\tilde{\varphi}_{L}(w,\bar{w})$ in the presence of the symmetric conformal boundary condition $A$.}
\label{TwistedBoundaryConfigurations} 
\end{figure}

Correlators involving bulk quasi-local operators living at the end of topological defect lines can be computed following a strategy similar to the one without topological defect lines; namely, we take the OPE of the bulk operators until we reduce everything to one-point functions in the presence of the boundary condition. An important conceptual ingredient is that we allow for topological defect lines to extend between a quasi-local operator and a topological endpoint at the boundary condition. A depiction of this configuration is shown on the left of Figure \ref{TwistedBoundaryConfigurations}, while a two-point configuration involving a genuine local operator and a quasi-local operator living at the end of a topological defect line is presented on the right.

To compute the one-point function of a quasi-local operator, one might be inclined to evaluate the inner product between the state associated with the quasi-local operator $\tilde{\varphi}_{L}(z,\bar{z})$ and the conventional boundary state Eqn. \eqref{UntwistedBoundaryState}, and then apply \eqref{untwistedonepointfunction}. However, it is readily evident that this always produces a vanishing result. 

Clearly, the vanishing result occurs because we incorrectly used the boundary state in the untwisted sector. Instead, we should have picked the twisted boundary state \eqref{TwistedBoundaryState}. In this situation, using manipulations similar to those in the preceding subsection, we obtain:
\begin{equation} \label{OnePointTwistedCorrelator}
    \langle \tilde{\varphi}_{L}(z, \bar{z}) \rangle_{A} =  \frac{\langle A_{L} | \tilde{\varphi}_{L} \rangle / \langle A | 0 \rangle}{|z - \bar{z}|^{2h_{\varphi}}}.
\end{equation}
Thus, the boundary state in the twisted sector determines the one-point function of quasi-local operators with a topological defect line ending at the symmetric boundary. See Appendix \ref{IsingExample} for a simple example of this construction in the Ising model. See also \cite{Fukusumi:2021zme}.

\section{Twisted BCFT of $N$ Dirac Fermions ($N$-DF)} \label{Section3}

In this section we apply the (twisted) BCFT formalism developed above to the theory of $N$ Dirac fermions. First, we review the standard story for the conformal boundary conditions we are interested in, after which we turn our attention to correlators. The key feature is that for the correlators to be non-vanishing, we must generically pair genuine local operators (fermion insertions) with non-genuine local operators. The latter carry charges outside the lattice supporting the untwisted boundary states, forcing the description of the Cardy state into a twisted sector. Thus, we construct the relevant boundary states in the twisted sector and use them to compute the two-point functions.

\subsection{Introduction to $N$-DF BCFT}

Let us briefly recall the setup in \cite{Smith:2019jnh}. The bulk theory is a fermionic theory of $N$ Dirac fermions. In the bulk, these fermions enjoy an $SO(2N)_L\times SO(2N)_R$ chiral symmetry. The underlying $U(1)^N \subset SO(2N)$ current algebra consists of left(right)-moving currents $J_i$ ($\bar{J}_i$), whose modes satisfy
\begin{equation}
    [J_{i,n}, J_{j,m}] = [\bar{J}_{i,n}, \bar{J}_{j,m}] = n\delta_{ij}\delta_{n+m,0}.
\end{equation}
The $U(1)^N$ primaries of the bulk Hilbert space are labeled as
\begin{equation} \label{TrivialVacuumSpectrum}
    J_{i,0}|\lambda, \bar{\lambda}\rangle = \lambda_i|\lambda, \bar{\lambda}\rangle, \quad \bar{J}_{i,0}|\lambda, \bar{\lambda}\rangle = \bar{\lambda}_i|\lambda, \bar{\lambda}\rangle, \quad \lambda,\bar{\lambda}\in\Z^N.
\end{equation}

We focus on boundaries that preserve a ${U(1)}^N$ subgroup of the chiral symmetry. The preserved $U(1)_\alpha$ symmetry has currents
\begin{equation}
    \mathcal{J}_{\alpha,n} = Q_{\alpha i}J_{i,n}, \quad \bar{\mathcal{J}}_{\alpha,n} = \bar{Q}_{\alpha i}\bar{J}_{i,n}, \quad \alpha = 1,\cdots,N.
\end{equation}
The boundary conditions are
\begin{equation} \label{QBoundaryCondition}
    (\mathcal{J}_{\alpha,n}+\bar{\mathcal{J}}_{\alpha,-n})|\mathcal{B}\rangle = 0,
\end{equation}
where $|\mathcal{B}\rangle$ is the boundary state. The $Q$ and $\bar{Q}$ matrices must satisfy the constraints
\begin{equation}
    Q_{\alpha i}Q_{\beta, i} = \bar{Q}_{\alpha i}\bar{Q}_{\beta, i}
\end{equation}
for the boundary conditions \eqref{QBoundaryCondition} to be imposed simultaneously.
Since the $U(1)_\alpha$ currents have a redundancy of $\mathcal{J}_\alpha \to U_{\alpha\beta}\mathcal{J}_\beta$, $\bar{\mathcal{J}}_\alpha \to U_{\alpha\beta}\bar{\mathcal{J}}_\beta$, we define a rational and orthogonal matrix
\begin{equation}
    \cR_{ij} = (\bar{Q}^{-1})_{i\alpha}Q_{\alpha j}
\end{equation}
that is invariant under the redundant transformation.
The boundary conditions \eqref{QBoundaryCondition} can then be written as
\begin{equation} \label{RBoundaryCondition}
    (\cR_{ij}J_{j,n}+\bar{J}_{i,-n})|\mathcal{B}\rangle = 0.
\end{equation}

Defining the lattice
\begin{equation} \label{UntwistedLattice}
    \Lambda[\cR] = \{\lambda \in \Z^N|\cR\lambda \in \Z^N\},
\end{equation}
we can find the Ishibashi states that solve \eqref{RBoundaryCondition}:
\begin{equation} \label{UntwistedIshibashiState}
    |\lambda, \cR\rrangle = \exp\left(-\sum_{n=1}^\infty\frac{1}{n}\cR_{ij}\bar{J}_{i,-n}J_{j,-n}\right)|\lambda, -\cR\lambda\rangle, \quad \lambda \in \Lambda[\cR].
\end{equation}
Then, we can expand the boundary state as
\begin{equation}
    |\mathcal{B} \rangle = \sum_{\lambda\in\Lambda[\cR]}a_\lambda|\lambda, \cR\rrangle.
\end{equation}
Notice that states with $\lambda$ such that $\cR \lambda \notin \mathbb{Z}^{N}$ do not contribute to the boundary state. In particular, this means that if we try to pair such a holomorphic local operator with charge $\lambda$ with any antiholomorphic local operator we will find a vanishing two-point function (recall Eqn. \eqref{untwistedonepointfunction}).

The coefficients $a_\lambda$ can be determined using the fact that, in the cluster decomposition limit:
\begin{equation} \label{ClusterDecomposition}
    \lim_{|z|, |\omega| \to 1^+}\frac{\langle\mathcal{O}_1(z)\mathcal{O}_2(\omega)\rangle_\mathcal{B}}{\langle\mathcal{O}_1(z)\rangle_\mathcal{B}\langle\mathcal{O}_2(\omega)\rangle_\mathcal{B}} = 1,
\end{equation}
where $\mathcal{O}_{1,2}(z)$ are local operators, and we have put the theory on the complement of the unit disk. Solving this condition as shown in Appendix B of \cite{Smith:2019jnh}, one can show that the boundary states are characterized by a periodic vector $\theta \in [0,2\pi)^N$
\begin{equation} \label{UntwistedCardyState}
    |\theta, \cR \rangle = a_0\sum_{\lambda\in\Lambda[\cR]}e^{i\gamma_\cR(\lambda)}e^{i\theta\cdot\lambda}|\lambda, \cR\rrangle,
\end{equation}
up to an unfixed prefactor $a_0$.
The normalization condition follows from the open-closed duality \eqref{CardyCondition}
\begin{equation}
    \Tr_{\cH_{\theta_1\theta_2}}(q^{L_0-c/24})
    = \langle \theta_2, \cR|\tilde{q}^{(L_0+\bar{L}_0-c/12)/2}|\theta_1, \cR \rangle.
\end{equation}
Plugging two of the boundary states \eqref{UntwistedCardyState} into the RHS, we get
\begin{equation} \label{openspectrum}
    \Tr_{\cH_{\theta_1\theta_2}}(q^{L_0-c/24}) = \frac{|a_0|^2}{\vol(\Lambda[\cR])}\sum_{\mu\in\Lambda^*[\cR]}\frac{q^{\frac{1}{2}\left(\mu -\frac{\theta_1-\theta_2}{2\pi}\right)^2}}{\eta(\tau)^N},
\end{equation}
where $\Lambda^*[\cR]$ is the dual lattice of $\Lambda[\cR]$. Requiring the overall factor on the RHS to be unity, we determine the prefactor of boundary states
\begin{equation} \label{ExactBoundaryState}
    |\theta, \cR \rangle = \sqrt{\vol({\Lambda[\cR]})}\sum_{\lambda\in\Lambda[\cR]}e^{i\gamma_\cR(\lambda)}e^{i\theta\cdot\lambda}|\lambda, \cR\rrangle.
\end{equation}

\subsection{Twisted Boundary in $N$-DF BCFT}

In the discussion of the Callan-Rubakov effect, we wish to consider two-point functions including a holomorphic fermion insertion. A holomorphic insertion carries a charge $\lambda$ that generically, however, is not in the lattice $\Lambda[\cR]$: $\lambda\notin\Lambda[\cR]$. Indeed, imposing the boundary condition \eqref{RBoundaryCondition} at the zeroth mode $n=0$, the charge of the antiholomorphic partner $\bar{\lambda}$ on the boundary has to be
\begin{equation} \label{FractionalChargeState}
    (\bar{\lambda}) = -\cR(\lambda) \notin \mathbb{Z}^N,
\end{equation}
and therefore the other insertion in the two-point function cannot generically correspond to a genuine local operator. 

Using the technology from Section \ref{TwistedBCFT}, however, it is clear how we can proceed to find a non-trivial correlator. In addition to the standard Cardy states, we can consider boundary states in a twisted sector. It will then be straightforward to pair a holomorphic fermion insertion with an antiholomorphic non-genuine local operator in such a way that the resulting boundary two-point function is non-vanishing.

In order to see this, let us consider the operator defined from the subgroup preserved by the boundary:
\begin{equation}
    \cD_i = e^{i2\pi[J_{i,0}+(Q^{-1})_{i\alpha}\bar{Q}_{\alpha j}\bar{J}_{j,0}]} = e^{i2\pi[J_{i,0}+(\cR^{-1})_{ij}\bar{J}_{j,0}]} = e^{i2\pi(J_{i,0}+\cR_{ji}\bar{J}_{j,0})},
\end{equation}
where the overall coefficient in the exponent has been chosen so that the operator acts trivially on the holomorphic sector of the untwisted Hilbert space $\mathcal{H}$. As we will see momentarily, this allows us to pair the holomorphic local operators with a non-trivial antiholomorphic non-genuine local counterpart in the corresponding twisted Hilbert space $\cH_{\cD_i}$. Note that $\cR$ is a rational matrix, and therefore the operator generates a $\mathbb{Z}_{\kappa_i}$ discrete symmetry such that
\begin{equation}
    (\cD_i)^{\kappa_i} = 1
\end{equation}
with $\kappa_i$ being the least common denominator of $\{\cR_{ji}\}_{1\leq j\leq N}$.

Modularity relates the action of the symmetry line over the untwisted Hilbert space to the spectrum of the twisted Hilbert space:
\begin{equation}
    \Tr_{\cH}(\cD_i q^{L_0-\frac{c}{24}}\bar{q}^{\bar{L}_0-\frac{c}{24}}) = \Tr_{\cH_{\cD_i}}(\tilde{q}^{L_0-\frac{c}{24}}\bar{\tilde{q}}^{\bar{L}_0-\frac{c}{24}}).
\end{equation}
Explicitly evaluating the spectrum of the twisted Hilbert space, we have:
\begin{equation} \label{twistedspectrum}
    \begin{split}
        &\Tr_{\cH}(\cD_i q^{L_0-\frac{c}{24}}\bar{q}^{\bar{L}_0-\frac{c}{24}}) \\
        &= \prod_{j=1}^N\Tr_{\cH_j}(q^{L_0-\frac{c}{24}})\Tr_{\cH_j}(\cD_i\bar{q}^{\bar{L}_0-\frac{c}{24}}) \\
        &= \prod_{j=1}^N\left(q^{-\frac{1}{48}}\prod_{r=1/2}^\infty(1+q^r)\right)^2\left(\bar{q}^{-\frac{1}{48}}\prod_{r=1/2}^\infty(1+e^{2\pi i\cR_{ji}}\bar{q}^r)\right)\left(\bar{q}^{-\frac{1}{48}}\prod_{r=1/2}^\infty(1+e^{-2\pi i\cR_{ji}}\bar{q}^r)\right) \\
        &= \prod_{j=1}^N\frac{1}{|\eta(\tau)|^2}\left(\sum_{n}q^{n^2/2}\right)\left(\sum_{m}e^{2\pi im\cR_{ji}}\bar{q}^{m^2/2}\right) \\
        &= \prod_{j=1}^N\frac{1}{|\eta(-1/\tau)|^2}\left(\sum_{n}\tilde{q}^{n^2/2}\right)\left(\sum_{m}\bar{\tilde{q}}^{(m-\cR_{ji})^2/2}\right).
    \end{split}
\end{equation}
Interpreting the spectrum as a collection of $U(1)^N \times U(1)^N$ primaries, the twisted Hilbert space associated with the defect line $\cD_i$ is
\begin{equation}
    \cH_{\cD_i} = \left\{|\lambda, \bar{\lambda}\rangle|\lambda\in\mathbb{Z}^N, (\bar{\lambda})_j\in\mathbb{Z}-\cR_{ji}\right\}.
\end{equation}
More generally, for a composite defect line $\cD_a=\prod_{1\leq i\leq N}(\cD_i)^{a_i}$, similar calculations give the twisted Hilbert space
\begin{equation} \label{GeneralTwistedHilbertSpace}
    \cH_{\cD_a} = \left\{|\lambda, \bar{\lambda}\rangle\big|\lambda\in\mathbb{Z}^N, \bar{\lambda}\in\mathbb{Z}^N-\cR a\right\}.
\end{equation}

Finally, we can identify the fractional charge state that arises from the boundary condition discussion \eqref{FractionalChargeState} to be in one of the twisted Hilbert spaces constructed above.
For a holomorphic state with charge $\lambda$, the boundary condition requires the system to have an antiholomorphic state carrying charge $\bar{\lambda}=-\cR\lambda$.
Comparing this with the general form \eqref{GeneralTwistedHilbertSpace}, we find that the antiholomorphic state lives in the twisted Hilbert space $\cH_{\cD_\lambda}$ with
\begin{equation} \label{AssociatedDefectLine}
    \cD_\lambda = \prod_{1\leq i\leq N}(\cD_i)^{(\lambda)_i} = e^{i2\pi\left[(\lambda)_iJ_{i,0}+\cR_{ji}(\bar{\lambda})_i\bar{J}_{j,0}\right]}.
\end{equation}


With the twisted Hilbert space $\cH_{\cD_a}$ constructed, we can discuss how to build boundary states in the twisted sector following Section \ref{TwistedBCFT}. From the boundary condition \eqref{RBoundaryCondition} at $n=0$, primaries $|\lambda,\bar{\lambda}\rangle\in\cH_{\cD_a}$ that can form Ishibashi states must be in the twisted lattice
\begin{equation}
    \lambda\in\Lambda_{\cD_a}[\cR] = \left\{\lambda\in\mathbb{Z}^N\big|-\cR\lambda\in\mathbb{Z}^N-\cR a\right\}, \quad
    \bar{\lambda}=-\cR\lambda.
\end{equation}
Comparing this with the lattice \eqref{UntwistedLattice} defined for the untwisted Hilbert space, we get a simple relation
\begin{equation}
    \lambda\in\Lambda[\cR] \quad \Leftrightarrow \quad
    \lambda+a\in\Lambda_{\cD_a}[\cR].
\end{equation}
That is, the twisted lattice is equivalent to the untwisted one up to a shift.

The boundary conditions \eqref{RBoundaryCondition} at $n\neq0$ can be solved by dressing the primaries with the same factor as in \eqref{UntwistedIshibashiState}. The twisted Ishibashi states then take the form
\begin{equation}
    |\lambda, \cR _{\cD_a} \rrangle = \exp\left(-\sum_{n=1}^\infty\frac{1}{n}\cR_{ij}\bar{J}_{i,-n}J_{j,-n}\right)|\lambda, -\cR\lambda\rangle, \quad \lambda \in \Lambda_{\cD_a}[\cR].
\end{equation}
To further compute the twisted Cardy states, we need to extend the cluster decomposition formula \eqref{ClusterDecomposition} to states and operators in twisted sectors.
The natural extension is
\begin{equation}
    \lim_{|z|, |\omega| \to 1^+}\frac{\langle\mathcal{O}(z)\mathcal{O}_{\cD_a}(\omega)\rangle_{\mathcal{B}}}{\langle\mathcal{O}(z)\rangle_\mathcal{B}\langle\mathcal{O}_{\cD_a}(\omega)\rangle_{\mathcal{B}}} = 1,
\end{equation}
where $\mathcal{O}_{\cD_a}$ and $|\mathcal{B}_{\cD_a} \rangle$ are a twisted operator and a twisted Cardy state with a defect line $\cD_a$ attached. With the generalized cluster decomposition, we can again fix the ratios of the Ishibashi state coefficients:
\begin{equation}
    \begin{split}
        |\theta, \cR_{\cD_a}  \rangle
        &= \sum_{\lambda\in\Lambda_{\cD_a}[\cR]}a_{\lambda, \cD_a}|\lambda, \cR _{\cD_a}\rrangle \\
        &= \bar{a}_0\sum_{\lambda\in\Lambda_{\cD_a}[\cR]}e^{i\gamma_\cR(\lambda)}e^{i\theta\cdot\lambda}|\lambda, \cR _{\cD_a} \rrangle.
    \end{split}
\end{equation}
Similar to the untwisted Cardy states \eqref{UntwistedCardyState}, twisted Cardy states are also parametrized by a periodic $N$-vector $\theta\in[0,2\pi)^N$.
To fix the overall coefficient $\bar{a}_0$, we plug twisted Cardy states into the twisted open-closed duality relation \eqref{twistedmodularity}, and get:
\begin{align}
    \begin{split}
        \quad {\langle \theta_2, \cR _{\cD_a}|}&q^{(L_0+\bar{L}_0-c/12)/2}|\theta_1, \cR_{\cD_a} \rangle \\
        &= |\bar{a}_0|^2 q^{-N/24} \sum_{\lambda\in\Lambda_{\cD_a}[\cR]}e^{i(\theta_1-\theta_2)\cdot\lambda}q^{\lambda^2/2}\prod_{n=1}^\infty\frac{1}{(1-q^n)^N} \\
        &= |\bar{a}_0|^2\sum_{\lambda\in\Lambda[\cR]}\frac{e^{i(\theta_1-\theta_2)\cdot(\lambda+a)+\pi i\tau(\lambda+a)^2}}{\eta(\tau)^N} \\
        &= |\bar{a}_0|^2\int d^Nx\sum_{\lambda\in\Lambda[\cR]}\frac{e^{i(\theta_1-\theta_2)\cdot(\lambda+a)+2\pi ix\cdot(\lambda+a)-\frac{\pi i}{\tau}x^2}}{\eta(-1/\tau)^N} \\
        &= \frac{|\bar{a}_0|^2}{\vol(\Lambda[\cR])}\int d^Nx\sum_{\mu\in\Lambda^*[\cR]}\delta^N\left(\mu-x-\frac{\theta_1-\theta_2}{2\pi}\right)\frac{e^{2\pi ia\cdot\left(x+\frac{\theta_1-\theta_2}{2\pi}\right)}\tilde{q}^{x^2/2}}{\eta(-1/\tau)^N} \\
        &= \frac{|\bar{a}_0|^2}{\vol(\Lambda[\cR])}\sum_{\mu\in\Lambda^*[\cR]}e^{2\pi ia\cdot\mu} \frac{\tilde{q}^{\frac{1}{2}\left(\mu - \frac{\theta_1-\theta_2}{2\pi}\right)^2}}{\eta(-1/\tau)^N} = \Tr_{\cH_{\theta_1\theta_2}}(\cD_a\tilde{q}^{L_0-c/24}).
    \end{split}
\end{align}
Therefore, matching with the spectrum \eqref{openspectrum} up to charges, we obtain:
\begin{equation} \label{twistedboundarystate}
    |\theta,\cR_{\cD_a} \rangle = \sqrt{\vol(\Lambda[\cR])}\sum_{\lambda\in\Lambda_{\cD_a}[\cR]}e^{i\gamma_\cR(\lambda)}e^{i\theta\cdot\lambda}|\lambda, \cR_{\cD_a} \rrangle,
\end{equation}
which has the same form as the Cardy state in the untwisted sector except that the summation is over different lattices. In addition, we find that the primaries in the open sector (labeled by $\mu \in \Lambda^*[\cR]$) carry charges $a\cdot\mu$ under $\cD_a$.

\subsubsection{Two-point Correlation Functions} \label{TwoPTCorrelators}

We can now construct two-point functions involving a holomorphic insertion of charge $\lambda$. The boundary condition then requires the antiholomorphic insertion to carry charge $\bar{\lambda}=-\cR\lambda$. However, since $\lambda$ does not generically belong to $\Lambda[\cR]$, two-point functions of two genuine local operators typically vanish. A non-trivial two-point function can nevertheless be obtained if the antiholomorphic insertion is a non-genuine local operator attached to a topological line defect. Studying such configurations requires boundary states in a twisted sector, such as those considered above.

It is instructive for the sake of comparison to first consider the case of $(\Box, \overline{\Box})$ boundary conditions where the calculation is standard. This boundary condition is defined by an $\cR$ matrix which is simply
\begin{equation}
    \cR_{ij} = -\delta_{ij},
\end{equation}
and the holomorphic and antiholomorphic insertions merely correspond to left-moving and right-moving fermions of the same charges. This is encoded in the two-point function:\footnote{Boundary conditions in the 2$d$ CFT problem are characterized by the $\theta$ angles written above. However, in the underlying four-dimensional problem we can set these angles to zero, $\theta = 0$, by field redefinitions and the Gauss law constraint. That is, a small but finite coupling prefers one boundary
condition, and there is no marginal deformation. See the discussion in Section~3.4 of~\cite{vanBeest:2023dbu}. For this reason, from now on we present our results for $\theta = 0$, although it is not difficult to put back the phases where necessary.}
\begin{equation} \label{boxboxbarcorrelator}
    \langle \psi_i(z) \tilde{\psi}^{\dagger}_{j}(\bar{w})\rangle_{\mathcal{B}} = \frac{\langle \mathcal{B} | (\psi_{i} \tilde{\psi}^{\dagger}_{j}) \rangle / \langle \mathcal{B} | 0 \rangle}{|z-\bar{w}|} = \frac{\delta_{ij}}{|z-\bar{w}|}.
\end{equation}

Meanwhile, the simplest context where the Callan-Rubakov effect takes place is for $(\Box, \Box)$ boundary conditions, defined by the $\cR$ matrix
\begin{equation}
    \cR_{ij} = \delta_{ij}-\frac{2}{N},
\end{equation}
for $1\leq i\leq N$. The holomorphic insertion is a left-moving fermion $\mathcal{O}_\lambda(z) = \psi_i(z)$. Let $\bar{\mathcal{O}}_{\bar{\lambda}}(\bar{w})$ be the non-genuine local primary operator attached to $\mathcal{D}_{\lambda}$ with $\bar{\lambda}=-\cR\lambda$ (recall the spectrum in \eqref{twistedspectrum} and the discussion below). Following Section \ref{TwistedBCFT}, the two-point function then becomes:
\begin{equation}
    \langle\psi_i(z)\bar{\mathcal{O}}_{\bar{\lambda}}(\bar{w})\rangle_{\mathcal{B}} = \frac{\langle \mathcal{B}_{\cD_\lambda} | (\psi_{i}\bar{\mathcal{O}}_{\bar{\lambda}}) \rangle / \langle \mathcal{B} | 0 \rangle}{|z-\bar{w}|} = \frac{1}{|z-\bar{w}|},
\end{equation}
where we have used the expression for the twisted boundary state \eqref{twistedboundarystate} and the untwisted one \eqref{ExactBoundaryState}, and the fact that the OPE of a holomorphic and an antiholomorphic operator gives simply the local operator with the corresponding holomorphic factorization.

\section{Unitarity from BCFT Correlators} \label{Section4}

The scattering of fermions off a monopole is effectively controlled by the s-wave sector. At leading order, higher partial waves are described by the bulk Dirac equation in the monopole background, whereas the s-wave reaches the monopole core and requires a boundary condition at the origin. In scattering language, this manifests as an apparent failure of unitarity if one keeps only the bulk monopole modes.

In this section, we first recall how this apparent failure of unitarity arises and how imposing the boundary condition at the origin resolves it \cite{PhysRevD.15.2287, Goldhaber:1977xw, Callias:1977cc, Callan:1983ed, Maldacena:1995pq}. We begin by recalling the case of $(\Box, \overline{\Box})$ boundary conditions, which can be imposed directly on the fermion fields. Here, unitarity can be recovered directly by examining the fields or, equivalently, by using the boundary condition obtained from BCFT. The more interesting case of the Callan-Rubakov effect corresponds to $(\Box, \Box)$ boundary conditions, which cannot be imposed directly on the fields. Proceeding by analogy with the first case, we instead compute the correlators using the twisted BCFT methods developed above and argue that unitarity is restored by comparing the resulting expressions.

The technical details of this section, such as the explicit form of the eigenfunctions and the partial-wave expansion, are summarized in Appendices \ref{ModeExpansionUnderMonopoleBackground} through \ref{AsymptoticBehaviorOfPlaneWaves}.

\subsection{Gauge-Covariant Asymptotic States and Partial-Wave Scattering} \label{Section4.1}

To discuss unitarity in the scattering of fermions in the monopole background, we follow the partial-wave analysis of \cite{Lee_2023}: set up asymptotic in- and out-states at spatial infinity $r\to\infty$, expand them in partial waves of the Dirac equation, and compare the incoming and outgoing partial-wave coefficients to read off the S-matrix. When the scattering does not involve long-range forces, the left/right-handed asymptotic states are momentum eigenstates that solve the free Dirac equation
\begin{equation}
    \psi_{L/R,\vec{p}}^{\text{plane}}(t,\vec r)=|p_{-/+})e^{i\vec p\cdot\vec r}, \qquad
    (\hat{p}\cdot\vec{\sigma})|p_\pm) = \pm|p_\pm),
\end{equation}
where $|p_\pm)$ are 3D spinors (see Appendix \ref{SpatialSpinors} for a more thorough discussion).
In the presence of a monopole, however, the vector potential does not decay fast enough at spatial infinity, and the asymptotic states must be modified to be compatible with the transition functions of the monopole bundle. For a minimal monopole, the vector potential is given by
\begin{equation}
    \vec{A}(\vec{r}) = \frac{1}{2r}\frac{\hat{r} \times \hat{n}}{1+\hat{r}\cdot\hat{n}}.
\end{equation}
The unit vector $\hat{n}$ corresponds to the patch choice of the vector potential.\footnote{
    For example, taking $\hat{n}=\hat{z}$ gives the vector potential in the northern patch, and the function is singular at the south pole $\hat{r}=-\hat{z}$.} The transition function that connects vector potentials in two different patches $\hat{n}$ and $\hat{n}'$ takes the form of a gauge transformation\footnote{We use the shorthand notation $|r_+)$ and $|r_+')$ to denote the spinors defined with respect to the reference vectors $\hat{n}$ and $\hat{n}'$.}
\begin{equation}
    \vec{A}'(\vec{r}) - \vec{A}(\vec{r}) = i\nabla(r_+'|r_+).
\end{equation}
Since the transition function does not vanish at spatial infinity, the asymptotic wave functions have to include extra factors to be compatible with the monopole. The gauge-covariant version of the asymptotic wave functions is then defined as
\begin{align}
    \psi_{L\vec p}^{\mathrm{plane,in}} &= |p_-)(p_-|r_+)e^{i\vec p\cdot\vec r},
    &
    \psi_{L\vec p}^{\mathrm{plane,out}} &= |p_-)(p_+|r_+)e^{i\vec p\cdot\vec r}, \\
    \psi_{R\vec p}^{\mathrm{plane,in}} &= |p_+)(p_-|r_+)e^{i\vec p\cdot\vec r},
    &
    \psi_{R\vec p}^{\mathrm{plane,out}} &= |p_+)(p_+|r_+)e^{i\vec p\cdot\vec r}.
\end{align}
Note that the in- and out-states now take different forms.\footnote{This is because the in-states should reproduce the free plane-wave states in the $\hat{r}=-\hat{p}$ direction, while the out-states should reproduce the free plane-wave states in the $\hat{r}=\hat{p}$ direction.}

We can expand the asymptotic states in the partial-wave basis labeled by the energy and the angular momentum quantum numbers $(E,\ell,m)$.\footnote{
    The angular momentum quantum numbers are defined with respect to the operators $J^2$ and $\hat{k}\cdot\vec{J}$ for some reference direction $\hat{k}$.
    Choosing $\hat{k}=\hat{z}$ corresponds to the standard choice of the spherical harmonics $Y_{\ell m}(\theta,\phi)$.
}
The results for the left-handed modes are 
\begin{equation} \label{mainInStateExpansionLeft}
     \psi_{L\vec{p}}^{\text{in}}(\vec{r}) = \pi\left[ \sqrt{2} \psi_{p00}^{(-\hat{p})}(\vec{r})+\sum_{\ell=1}^\infty\sqrt{2(2\ell+1)}e^{-\frac{i\pi n}{2}}\psi_{p\ell0}^{(-\hat{p})}(\vec{r})\right],
\end{equation}
\begin{equation} \label{mainOutStateExpansionLeft}
    \psi_{L\vec{p}}^{\text{out}}(\vec{r}) = \pi\sum_{\ell=1}^\infty\sqrt{2(2\ell+1)}e^{\frac{i\pi n}{2}}\psi_{p\ell(-1)}^{(\hat{p})}(\vec{r}),
\end{equation}
with $n = \sqrt{\ell(\ell+1)}$ for $\ell \geq 1$. Note that the s-wave mode $\ell=0$ in the partial-wave basis appears in the expansion of the in-state, but not of the out-state. For the right-handed fermions, we have:
\begin{equation}
    \psi_{R\vec{p}}^{\text{in}}(\vec{r}) = \pi\sum_{\ell=1}^\infty\sqrt{2(2\ell+1)}e^{\frac{-i\pi n}{2}}\psi_{(-p)\ell(-1)}^{(-\hat{p})}(\vec{r}),
\end{equation}
\begin{equation} \label{mainOutStateExpansionRight}
     \psi_{R\vec{p}}^{\text{out}}(\vec{r}) = \pi\left[ \sqrt{2} \psi_{(-p)00}^{(\hat{p})}(\vec{r})+\sum_{\ell=1}^\infty\sqrt{2(2\ell+1)}e^{\frac{i\pi n}{2}}\psi_{(-p)\ell0}^{(\hat{p})}(\vec{r})\right],
\end{equation}
where the s-wave mode is in the out-state but not in the in-state. In terms of the creation and annihilation operators, the expansions above can be written as
\begin{equation} \label{eq:operator_relation1}
    \hat{c}_{L\vec{p}}^{\text{in}\dagger} = \pi\left[ \sqrt{2} \hat{c}_{Lp00}^{(-\hat{p})\dagger}+\sum_{\ell=1}^\infty\sqrt{2(2\ell+1)}e^{-\frac{i\pi n}{2}}\hat{c}_{Lp\ell0}^{(-\hat{p})\dagger}\right],
\end{equation}
\begin{equation}
    \hat{c}_{L\vec{p}}^{\text{out}\dagger} = \pi\sum_{\ell=1}^\infty\sqrt{2(2\ell+1)}e^{\frac{i\pi n}{2}}\hat{c}_{Lp\ell(-1)}^{(\hat{p})\dagger},
\end{equation}
\begin{equation}
    \hat{c}_{R\vec{p}}^{\text{in}\dagger} = \pi\sum_{\ell=1}^\infty\sqrt{2(2\ell+1)}e^{\frac{-i\pi n}{2}}\hat{c}_{Rp\ell(-1)}^{(-\hat{p})\dagger},
\end{equation}
\begin{equation} \label{eq:operator_relation4}
     \hat{c}_{R\vec{p}}^{\text{out}\dagger} = \pi\left[ \sqrt{2} \hat{c}_{Rp00}^{(\hat{p})\dagger}+\sum_{\ell=1}^\infty\sqrt{2(2\ell+1)}e^{\frac{i\pi n}{2}}\hat{c}_{Rp\ell0}^{(\hat{p})\dagger}\right].
\end{equation}
On the left-hand side, the operators in the momentum basis only act on asymptotic states, like the asymptotic operators that appear in the derivation of the LSZ formula.
Meanwhile, the operators in the partial-wave basis can be used on all time slices as the partial-wave functions are exact solutions to the Dirac equation.

The one-to-one S-matrix can now be calculated straightforwardly. For instance, the left-to-left S-matrix from momentum $\vec{p}$ to momentum $\vec{q}$ is given by
\begin{equation} \label{LtoLSMatrix}
    \begin{split}
        &\mathcal{M}_{L\to L}(\vec{p},\vec{q})
        = \langle\vec{q},L|\vec{p},L\rangle = \langle0|\hat{c}_{L\vec{q}}^{\text{out}}\hat{c}_{L\vec{p}}^{\text{in}\dagger}|0\rangle \\ 
        &\quad = \pi^2\langle0|\left[\sum_{\ell=1}^\infty\sqrt{2(2\ell+1)}e^{\frac{i\pi n}{2}}\hat{c}_{Lq\ell(-1)}^{(\hat{q})\dagger}\right]^\dagger\left[2\hat{c}_{Lp00}^{(-\hat{p})\dagger}+\sum_{\ell=1}^\infty\sqrt{2(2\ell+1)}e^{-\frac{i\pi n}{2}}\hat{c}_{Lp\ell0}^{(-\hat{p})\dagger}\right]|0\rangle.
    \end{split}
\end{equation}
With the monopole harmonics relation \eqref{SpinWeightedKChanged} that connects harmonics of different spin directions $\hat{k}$, we can read off the corresponding expression for the creation (annihilation) operators:
\begin{equation} \label{OperatorKChanged}
    \hat{c}_{L q \ell m}^{(\hat{q})\dagger} = \sqrt{\frac{4\pi}{2\ell+1}}\sum_{m'}Y_{\ell m}^{-m'}(-\hat{p},\hat{q}) \, \hat{c}_{L q \ell m'}^{(-\hat{p})\dagger}.
\end{equation}

Plugging \eqref{OperatorKChanged} into \eqref{LtoLSMatrix} and applying anticommutation relations in \eqref{AnticommutationL}, the result is
\begin{equation}
    \mathcal{M}_{L\to L}(\vec{p},\vec{q}) = \frac{4\pi^2\delta(p-q)}{p^2}\sum_{\ell=1}^\infty\sqrt{\pi(2\ell+1)}e^{-i\pi n} Y^{1}_{\ell 0}(\hat{q},-\hat{p}).
\end{equation}
The $\delta$ function on the right-hand side implies energy conservation, which is a universal factor of  S-matrices. As usual, we define the reduced S-matrix $\mathcal{S}(\vec{p},\vec{q})$ as the remaining piece:
\begin{equation}
    \mathcal{M}(\vec{p},\vec{q}) = \frac{4 \pi^{2} \delta(p-q)}{p^{2}} \mathcal{S}(\vec{p},\vec{q}).
\end{equation}
Then, we have that:
\begin{equation} \label{mainReducedSLtoL}
    \mathcal{S}_{L\to L}(\hat{p},\hat{q}) = \sum_{\ell=1}^\infty\sqrt{\pi(2\ell+1)}e^{-i\pi n} Y^{1}_{\ell 0}(\hat{q},-\hat{p}),
\end{equation}
and the reduced S-matrices of other processes are
\begin{align} \label{mainReducedSRtoR}
    \mathcal{S}_{R\to R}(\hat{p},\hat{q}) &= \sum_{\ell=1}^\infty\sqrt{\pi(2\ell+1)}e^{i\pi n}Y^{0}_{\ell (-1)}(\hat{q},\hat{p}), \\[0.2cm] \mathcal{S}_{L\to R}(\hat{p},\hat{q}) &= \mathcal{S}_{R\to L}(\hat{p},\hat{q}) = 0.
\end{align}
Since the bulk equation of motion preserves helicity and the boundary condition at the monopole core has not been imposed (yet), there are no non-trivial helicity-flipped scattering processes.

For unitary scattering, the reduced S-matrix must satisfy the unitarity condition
\begin{equation}
    \int\frac{d^2\hat{q}}{(2\pi)^2} \ \mathcal{S}^\dagger(\hat{q}, \hat{p}')\mathcal{S}(\hat{q}, \hat{p}) = \delta^{(2)}(\hat{p}'-\hat{p}) = \frac{1}{2\pi}\delta(1-\hat{p}\cdot\hat{p}').
\end{equation}
For a right-handed in-state, \eqref{mainReducedSRtoR} indeed saturates the unitarity condition:
\begin{equation}
    \int\frac{d^2\hat{q}}{(2\pi)^2} \ \mathcal{S}_{R\to R}^\dagger(\hat{q}, \hat{p}')\mathcal{S}_{R\to R}(\hat{q}, \hat{p}) = \sum_{j=1}^\infty\sqrt{\frac{2j+1}{4\pi}}Y_{j(-1)}^1(-\hat{p}', -\hat{p}) = \delta^{(2)}(\hat{p}'-\hat{p}),
\end{equation}
where \eqref{SpinWeightedIdentity} has been used to express the spin-weighted spherical harmonics as a delta function. If one proceeds in an analogous manner for a left-handed in-state, however, one finds that \eqref{mainReducedSLtoL} violates unitarity:
\begin{align} \label{mainLeftHandedViolateUnitarity}
    \int\frac{d^2\hat{q}}{(2\pi)^2} \ \mathcal{S}_{L\to L}^\dagger(\hat{q}, \hat{p}')\mathcal{S}_{L\to L}(\hat{q}, \hat{p}) &= \sum_{j=1}^\infty\sqrt{\frac{2j+1}{4\pi}}Y_{j0}^0(-\hat{p}', -\hat{p}) \nonumber \\ &= \delta^{(2)}(\hat{p}'-\hat{p})-\frac{1}{\sqrt{4\pi}}Y_{00}^0(-\hat{p}', -\hat{p}).
\end{align}
Tracing the missing piece back to the relations \eqref{mainInStateExpansionLeft} and \eqref{mainOutStateExpansionLeft}, one can see that the missing piece comes from the absence of the s-wave mode in the out-state. The bulk monopole partial-wave analysis thus identifies exactly which channel must be completed by the physics at the monopole core.

\subsection{S-Wave Boundary Condition and BCFT}

To isolate the s-wave sector, we decompose the four-dimensional Dirac fermion into partial waves as
\begin{equation} \label{partialWaveDecomposition}
    \Psi(t, \vec{r}) = \begin{pmatrix}
        \Psi_L(t, \vec{r})\\
        \Psi_R(t, \vec{r})
    \end{pmatrix}
    = \frac{1}{2 \sqrt{2} \pi r}\begin{pmatrix}
        \chi_L(t,r)|r_+)\\
        i\chi_R^\dagger(t,r)|r_+)
    \end{pmatrix}
    +(\text{higher partial waves}).
\end{equation}
The fermionic fields $\chi_L(t,r)$ and $\chi_R(t,r)$ live on the half-line $r\geq0$. Away from the monopole core the bulk theory is a free fermion, and the two chiral fields propagate independently. The missing piece in the scattering problem is a boundary condition at $r=0$ that relates the left- and right-handed s-wave modes.

The two-dimensional fields can be recovered from the four-dimensional field by the following angular projections:
\begin{align}
    \chi_L(t,r) &= 2\sqrt{2} \pi r \int \frac{d^2\hat{r}}{4\pi}\,(r_+|\Psi_L(t,\vec{r}), \label{twodimfieldfromfourdimfield1} \\[0.1cm] \chi_R(t,r) &= 2\sqrt{2} \pi r i \int \frac{d^2\hat{r}}{4\pi}\,\Psi_R^\dagger(t,\vec{r})|r_+). \label{twodimfieldfromfourdimfield2}
\end{align}
Thus, the two-dimensional fields can be related to the creation (annihilation) operators of the four-dimensional fields:
\begin{align}
    \chi_L(t,r)
    &=\int_0^\infty E\,dE\left[
        \hat{c}_{LE00}^{(\hat{k})}e^{-iE(t+r)}
        +\hat{d}_{LE00}^{(\hat{k})\dagger}e^{iE(t+r)}
    \right], \\
    \chi_R(t,r)
    &= i \int_0^\infty E\,dE\left[
        \hat{c}_{RE00}^{(\hat{k})\dagger}e^{iE(t-r)}
        +\hat{d}_{RE00}^{(\hat{k})}e^{-iE(t-r)}
    \right],
\end{align}
which follows from substituting the s-wave functions \eqref{PositiveEnergyMode2} and \eqref{NegativeEnergyMode2} into \eqref{leftfield} and \eqref{rightfield}.

A boundary condition by definition provides an overlap between the left- and right-handed movers. Time-translation invariance makes this overlap diagonal in energy, while the normalization of the eigenfunctions (see \eqref{normalization}) fixes its energy dependence. We therefore write
\begin{equation} \label{moderelation}
    \langle0|\hat{c}_{LE00}^{(\hat{k})}
    \hat{c}_{RE'00}^{(\hat{k})\dagger}|0\rangle_{\mathcal{B}}
    = \mathcal{C}^{\chi}_{\mathcal{B}} \frac{\delta(E-E')}{E^2}
\end{equation}
for some constant $\mathcal{C}^{\chi}_{\mathcal{B}}$ defined through the boundary condition.

To see how the boundary condition imposed as a condition over the modes through \eqref{moderelation} relates to the boundary condition defined through BCFT methods, let us compute the two-point function between $\chi_L(t,r)$ and $\chi_R(t',r')$ using \eqref{twodimfieldfromfourdimfield1}-\eqref{twodimfieldfromfourdimfield2} and \eqref{moderelation}. Including the $i\epsilon$ prescription for the time-ordered correlator to match the Euclidean BCFT correlator and defining $\Delta=(t-t')+(r+r')$, we have:
\begin{align} 
        \hspace{-0.5cm} \langle\chi_L(t,r)\chi_R(t',r')\rangle_{\mathcal{B}} &= \left\langle
        \left[r\int\frac{d^2\hat{r}}{4\pi}\,(r_+|\Psi_L(t,\vec{r})\right]
        \left[ir'\int\frac{d^2\hat{r}'}{4\pi}\,\Psi_R^\dagger(t',\vec{r}')|r'_+)\right]
        \right\rangle_{\mathcal{B}} \nonumber \\
        &= i\int_0^\infty E\,dE\int_0^\infty E'\,dE'\,
        e^{-iE(t+r)}e^{iE'(t'-r')}
        \langle0|\hat{c}_{LE00}^{(\hat{k})}
        \hat{c}_{RE'00}^{(\hat{k})\dagger}|0\rangle_{\mathcal{B}} \nonumber \\
        &= i\int_0^\infty E\,dE\int_0^\infty E'\,dE'\,
        e^{-iE(t+r)}e^{iE'(t'-r')}
        \mathcal{C}^{\chi}_{\mathcal{B}} \frac{\delta(E-E')}{E^2} \nonumber \\
        &= i\int_0^\infty dE\, \mathcal{C}^{\chi}_{\mathcal{B}} \,
        e^{-iE(\Delta-i\epsilon)} = \frac{\mathcal{C}^{\chi}_{\mathcal{B}}}{\Delta-i\epsilon}. \label{mainTwoFormulationRelationMerged}
\end{align}
As expected, the correlator computed from the boundary condition defined via the creation (annihilation) operators \eqref{moderelation} reproduces the BCFT correlator. Moreover, the coefficient in \eqref{moderelation} can be identified with the boundary state overlap coefficient in \eqref{boxboxbarcorrelator}.

More importantly, the equation above relates the BCFT correlator to a corresponding expression for the creation and annihilation operators of the four-dimensional fields. In this sense, the BCFT correlator directly corresponds to the relevant $S$-matrix element, and it is clear that the restoration of unitarity can be expressed as a condition on the coefficient of
the BCFT correlator. In the specific context of the free fermion theory (with $(\square,\bar{\square})$ boundary conditions), this coefficient is simply unity (recall Eqn.~\eqref{boxboxbarcorrelator}). As we will see, this is indeed the appropriate coefficient to restore unitarity~\cite{PhysRevD.15.2287, Goldhaber:1977xw, Callias:1977cc, Callan:1983ed, Maldacena:1995pq}.

All in all, the mixed overlap of the four-dimensional s-wave operators is
\begin{equation}
    \langle0|\hat{c}_{LE00}^{(\hat{k})}
    \hat{c}_{RE'00}^{(\hat{k})\dagger}|0\rangle_{\mathcal{B}}
    = \frac{\delta(E-E')}{E^2}.
\end{equation}

Since the operators of different helicities are not independent, we can introduce a single operator $\hat{c}_{E00}^\dagger$ that creates the s-wave mode of both helicities at the monopole core. Comparing with the two-point function:
\begin{equation} \label{mainRelatedSWaveBasisMerged1}
    \hat{c}_{E00}
    = \hat{c}_{LE00}^{(\hat{k})}
    = \hat{c}_{RE00}^{(\hat{k})}.
\end{equation}
Similarly, the antiparticle annihilation operators of the left- and right-handed s-wave modes are related by
\begin{equation} \label{mainRelatedSWaveBasisMerged2}
    \hat{d}_{E00}
    = \hat{d}_{LE00}^{(\hat{k})}
    = \hat{d}_{RE00}^{(\hat{k})}.
\end{equation}

The relations \eqref{mainRelatedSWaveBasisMerged1} and \eqref{mainRelatedSWaveBasisMerged2} affect only the s-wave terms in the asymptotic-wave expansions. It is thus straightforward to check that the helicity-preserving reduced S-matrices are unchanged, while the boundary condition generates an extra helicity-flipped contribution:
\begin{equation}
    \mathcal{S}_{L\to R}(\hat{p},\hat{q})
    =\frac{1}{2}.
\end{equation}
Its contribution to the unitarity relation is exactly the missing s-wave projector:
\begin{equation}
    \int\frac{d^2\hat{q}}{(2\pi)^2}\,
    \mathcal{S}_{L\to R}^\dagger(\hat{q},\hat{p}')
    \mathcal{S}_{L\to R}(\hat{q},\hat{p})
    =\frac{1}{4\pi}
    =\frac{1}{\sqrt{4\pi}}Y_{00}^0(-\hat{p}',-\hat{p}).
\end{equation}
Combining this result with \eqref{mainLeftHandedViolateUnitarity}, we obtain
\begin{equation}
    \int\frac{d^2\hat{q}}{(2\pi)^2}\left[
    \mathcal{S}_{L\to L}^\dagger(\hat{q},\hat{p}')
    \mathcal{S}_{L\to L}(\hat{q},\hat{p})
    +\mathcal{S}_{L\to R}^\dagger(\hat{q},\hat{p}')
    \mathcal{S}_{L\to R}(\hat{q},\hat{p})
    \right]
    =\delta^{(2)}(\hat{p}'-\hat{p}),
\end{equation}
so the boundary condition supplies precisely the contribution required by unitarity.

It is straightforward to generalize the calculation above to the case of $N$ flavors of fermions, with both chiralities transforming in the fundamental representation of $SU(N)$:
\begin{equation}
    \Psi_L\to(\Psi_L)_i, \qquad \Psi_R\to(\Psi_R)_i,
    \qquad i=1,\ldots,N.
\end{equation}
This is the $(\square,\bar{\square})$ case. Since the bulk action contains no flavor-mixing terms, different flavors propagate independently away from the monopole core. The higher partial-wave operators therefore acquire a flavor index without otherwise changing relative to the prior calculation: $\hat{c}_{LE\ell m}^{(\hat{k})}\to\hat{c}_{LE\ell m i}^{(\hat{k})}, \quad
    \hat{c}_{RE\ell m}^{(\hat{k})}\to\hat{c}_{RE\ell m i}^{(\hat{k})}$, and the monopole core relates opposite helicities with the same flavor index:
\begin{equation}
    \hat{c}_{E00i}
    =\hat{c}_{LE00i}^{(\hat{k})}
    =\hat{c}_{RE00i}^{(\hat{k})}.
\end{equation}
The corresponding reduced S-matrices are therefore:
\begin{equation}
    \left[\mathcal{S}_{L\to L}(\hat{p},\hat{q})\right]_{i}^{j}
    =\delta_{i}^{j}\sum_{\ell=1}^\infty\sqrt{\pi(2\ell+1)}e^{-i\pi n}Y_{\ell0}^{1}(\hat{q},-\hat{p}),
\end{equation}
\begin{equation}
    \left[\mathcal{S}_{R\to R}(\hat{p},\hat{q})\right]_{i}^{j}
    = \delta_{i}^{j} \sum_{\ell=1}^\infty\sqrt{\pi(2\ell+1)}e^{-i\pi n}Y_{\ell (-1)}^{0}(\hat{q}, -\hat{p}),
\end{equation}
\begin{equation}
    \left[\mathcal{S}_{L\to R,\mathcal{B}_\theta}(\hat{p},\hat{q})\right]_{i}^{j}
    = \delta_{i}^{j} \frac{1}{2}, \qquad
    \left[\mathcal{S}_{R\to L}(\hat{p},\hat{q})\right]_{i}^{j} = 0.
\end{equation}

Finally, let us comment on the more interesting case where the left- and right-handed fermions transform in conjugate representations of the flavor group. This is the case of the $(\square,\square)$ boundary conditions relevant to the Callan-Rubakov effect. In this case, the BCFT correlator contains a non-diagonal flavor structure $\mathcal{R}$, as discussed in Section~\ref{TwoPTCorrelators}. In this scenario, the picture relating the left- and right-handed s-wave operators is not so transparent since there is no clear four-dimensional field-theoretic description for the outgoing states.

Nevertheless, since the BCFT correlators take the same value as in \eqref{boxboxbarcorrelator}, one may leverage Eqn. \eqref{mainTwoFormulationRelationMerged}, reverse-engineer it, and propose that the BCFT correlators define a corresponding expression in terms of modes, or equivalently, the numerical value for the corresponding channel in the scattering S-matrix. This determines the helicity-flipped S-matrix:
\begin{equation}
    S_{\psi_{i} \to \bar{\mathcal{O}}_{\bar{\lambda}}}(\hat{p},\hat{q}) = \frac{1}{2},
\end{equation}
and since the s-wave fermion still scatters into only one channel by our discussion of the boundary condition, we recover unitarity just as in the discussion above. We just replace the standard right-handed fermion by the operator on the twisted sector. Otherwise, the numerical contribution remains the same and we have: 
\begin{equation}
    \int\frac{d^2\hat{q}}{(2\pi)^2}\left[
    \mathcal{S}_{L\to L}^\dagger(\hat{q},\hat{p}')
    \mathcal{S}_{L\to L}(\hat{q},\hat{p})
    + S_{\psi_{i} \to \bar{\mathcal{O}}_{\bar{\lambda}}}^\dagger(\hat{q},\hat{p}')
    S_{\psi_{i} \to \bar{\mathcal{O}}_{\bar{\lambda}}}(\hat{q},\hat{p})
    \right]
    =\delta^{(2)}(\hat{p}'-\hat{p}),
\end{equation}
which is indeed the statement that unitarity is reproduced.

\acknowledgments
We thank Z. Komargodski, Y. Wang, S-H. Shao, K. Ohmori, D. Tong, and H-T. Lam. for valuable conversations. LLMs (Claude and ChatGPT) were used for proofreading and editing text. DGS is supported by the Society of Fellows at Harvard University and the US Department of Energy under grant DE-SC0007870. CC, DGS acknowledge support from the Simons Collaboration on Global Categorical Symmetries, the US Department of Energy Grant 5-29073, and the Sloan Foundation. LTW is supported by the Department of Energy grant DE-SC0009924.

\appendix

\section{Mode Expansion under Monopole Background} \label{ModeExpansionUnderMonopoleBackground}

When discussing mode expansion of fields, one needs to find a basis of eigenfunctions that solves the equation of motion of the field.
In our case, the equation of motion is the Dirac equation
\begin{equation}
    0 = \gamma^\mu(-i\partial_\mu+A_\mu)\Psi(t,r)
\end{equation}
with a monopole gauge potential $A_\mu = (A_0,\vec{A})$
\begin{equation} \label{MonopoleBackground}
    \vec{A}(\vec{r}) = \frac{1}{2r}\frac{\vec{r}\times\hat{n}}{r+\vec{r}\cdot\hat{n}}, \quad r = |\vec{r}|.
\end{equation}
In this potential, the auxiliary unit vector $\hat{n}$ indicates which coordinate patch we use.\footnote{
    For example, taking $\hat{n}=\hat{z}$ corresponds to choosing the patch on the north pole. To study the gauge potential at the south pole, we change the patch, considering the corresponding gauge transition function. For more details, see Appendix \ref{SpatialSpinors}.
}
After taking the Weyl representation for the Gamma matrices and separating the time dependence of the field from its spatial part
\begin{equation}
    \Psi_{E}(t, \vec{r}) = e^{-iEt}\begin{pmatrix}
        \psi_L(\vec{r})\\
        \psi_R(\vec{r})
    \end{pmatrix},
\end{equation}
we get the chiral equation of motion for fixed energy modes
\begin{equation}
    E\psi_L(\vec{r}) = \hat{H}[\psi_L](\vec{r}) = -\vec{\sigma}\cdot\left[-i\nabla+\vec{A}(\vec{r})\right]\psi_L(\vec{r}),
\end{equation}
\begin{equation}
    E\psi_R(\vec{r}) = \hat{H}[\psi_R](\vec{r}) = \vec{\sigma}\cdot\left[-i\nabla+\vec{A}(\vec{r})\right]\psi_R(\vec{r}).
\end{equation}
As in the usual partial-wave analysis, in addition to energy, we also label the solutions with respect to the total angular momentum $\vec{J}^2$ and its component $\vec{J}\cdot\hat{k}$ in some specific direction $\hat{k}$.\footnote{ The most common choice is $\hat{k}=\hat{z}$ with operators $J^2$ and $J_z$.} The eigenfunction equations are
\begin{equation} 
    \vec{J}^2[\psi](\vec{r}) = \ell(\ell+1)\psi(\vec{r}), \quad
    (\vec{J}\cdot\hat{k})[\psi](\vec{r}) = m\psi(\vec{r}),
\end{equation}
and the corresponding normalization condition is
\begin{equation} \label{normalization}
    \langle {E\ell m}_{(\hat{k})}|{E'\ell'm'}_{(\hat{k})}\rangle = \int d^3\vec{r} \ \psi_{E\ell m}^{(\hat{k})\dagger}\psi_{E'\ell'm'}^{(\hat{k})} = \frac{\delta(E-E')}{E^2}\delta_{\ell\ell'}\delta_{mm'}.
\end{equation}

The complete set of normalized solutions labeled by $(E,\ell,m)$ for positive energy $E > 0$ and $\ell \geq 1$ is then given by \cite{PhysRevD.15.2287,Lee_2023}:
\begin{equation} \label{PositiveEnergyMode1}
    \psi_{E\ell m}^{(\hat{k})}(\vec{r}) = \frac{1}{\sqrt{2\pi}}\left\{\left[j_n(Er)+ij_{n-1}(Er)\right]|r_+)Y_{\ell m}^0(\hat{r}, \hat{k})+\left[j_n(Er)-ij_{n-1}(Er)\right]|r_-)Y_{\ell m}^1(\hat{r}, \hat{k})\right\},
\end{equation}
where $n = \sqrt{\ell(\ell+1)}$, $m$ runs from $-\ell$ to $\ell$, $j_\nu(x)$ is the spherical Bessel function of the first kind, $Y_{\ell m}^s(\hat{r}, \hat{k})$ is the spin-weighted spherical harmonics, and $|r_\pm)$ are two-component spatial spinors (see Appendix \ref{SpatialSpinors} for more information on spatial spinors and spin-weighted spherical harmonics). For negative energy modes $(-E)\leq0$, the corresponding basis is
\begin{equation} \label{NegativeEnergyMode1}
    \psi_{(-E)\ell m}^{(\hat{k})}(\vec{r}) = \frac{1}{\sqrt{2\pi}}\left\{\left[j_n(Er)-ij_{n-1}(Er)\right]|r_+)Y_{\ell m}^0(\hat{r}, \hat{k})+\left[j_n(Er)+ij_{n-1}(Er)\right]|r_-)Y_{\ell m}^1(\hat{r}, \hat{k})\right\}.
\end{equation}

For the s-wave $\ell = 0$, however, we find that the corresponding solution
\begin{equation} \label{PositiveEnergyMode2}
    \psi_{E00}^{(\hat{k})}(\vec{r}) \propto \frac{e^{-iEr}}{Er}|r_+)Y_{00}^0(\hat{r}, \hat{k}),
\end{equation}
cannot be properly normalized according to \eqref{normalization}, essentially because the mode comes unpaired and thus the self-inner-product cannot be made real as in the RHS of \eqref{normalization}.\footnote{This is a manifestation of the fact that the monopole Dirac Hamiltonian is not hermitian until we impose the boundary condition.} The issue is resolved by imposing the boundary condition at $r = 0$. This can be made explicit for the $(\Box, \overline{\Box})$ boundary condition, in which case we write the corresponding spatial eigenfunction as 
\begin{equation}
    \Psi^{(\hat{k})}_{E00}(t, \vec{r}) = \frac{A_{s}}{Er}\begin{pmatrix}
        e^{-iEr} |r_+)Y_{00}^0(\hat{r},\hat{k}) \\
        e^{i \theta} e^{iEr} |r_+)Y_{00}^0(\hat{r},\hat{k})
    \end{pmatrix},
\end{equation}
where $A_{s}$ is an overall normalization constant not determined by the boundary condition. The corresponding wavefunction can now describe a proper orthonormal state normalized according to \eqref{normalization}, and we find $A_{s} = 1/\sqrt{2\pi}$. Splitting into incoming and outgoing modes:
\begin{equation} \label{PositiveEnergyMode2}
    \psi_{E00}^{(\hat{k})}(\vec{r}) = \frac{1}{\sqrt{2\pi}}\frac{e^{-iEr}}{Er}|r_+)Y_{00}^0(\hat{r}, \hat{k}),
\end{equation}
\begin{equation} \label{NegativeEnergyMode2}
    \psi_{(-E)00}^{(\hat{k})}(\vec{r}) = \frac{1}{\sqrt{2\pi}}\frac{e^{iEr}}{Er}|r_+)Y_{00}^0(\hat{r}, \hat{k}).
\end{equation}

In the $(E,\ell,m)$ basis, the fermion field in the Heisenberg picture has the form
\begin{equation} \label{leftfield}
    \hat{\Psi}_L(t, \vec{r}) = \sum_{\ell, m}\int_0^\infty E^2 dE \ \left[\hat{c}_{LE\ell m}^{(\hat{k})}e^{-iEt}\psi_{E\ell m}^{(\hat{k})}+\hat{d}_{LE\ell m}^{(\hat{k})\dagger} e^{iEt}\psi_{(-E)\ell m}^{(\hat{k})}\right],
\end{equation}
\begin{equation} \label{rightfield}
    \hat{\Psi}_R(t, \vec{r}) = \sum_{\ell, m}\int_0^\infty E^2 dE \ \left[\hat{c}_{RE\ell m}^{(\hat{k})}e^{-iEt}\psi_{(-E)\ell m}^{(\hat{k})}+\hat{d}_{RE\ell m}^{(\hat{k})\dagger} e^{iEt}\psi_{E\ell m}^{(\hat{k})}\right].
\end{equation}
Applying the equal-time canonical quantization condition
\begin{equation}
    \left\{\hat{\Psi}_L(t,\vec{r}), \hat{\Psi}_L^\dagger(t,\vec{r}')\right\} = \left\{\hat{\Psi}_R(t,\vec{r}), \hat{\Psi}_R^\dagger(t,\vec{r}')\right\} = \delta^{(3)}(\vec{r}-\vec{r}'),
\end{equation}
the anticommutation relations are
\begin{equation} \label{AnticommutationL}
    \left\{\hat{c}_{LE\ell m}^{(\hat{k})},\hat{c}_{LE'\ell'm'}^{(\hat{k})\dagger}\right\} = \left\{\hat{d}_{LE\ell m}^{(\hat{k})},\hat{d}_{LE'\ell'm'}^{(\hat{k})\dagger}\right\} = \frac{\delta(E-E')}{E^2}\delta_{\ell\ell'}\delta_{mm'},
\end{equation}
\begin{equation} \label{AnticommutationR}
    \left\{\hat{c}_{RE\ell m}^{(\hat{k})},\hat{c}_{RE'\ell'm'}^{(\hat{k})\dagger}\right\} = \left\{\hat{d}_{RE\ell m}^{(\hat{k})},\hat{d}_{RE'\ell'm'}^{(\hat{k})\dagger}\right\} = \frac{\delta(E-E')}{E^2}\delta_{\ell\ell'}\delta_{mm'},
\end{equation}
with the s-wave modes identified as in \eqref{mainRelatedSWaveBasisMerged1} and \eqref{mainRelatedSWaveBasisMerged2} as imposed by the boundary condition.

\section{Gauge-Covariant Asymptotic States}
\label{AsymptoticStatesAndScatteringAmplitudes}

When it comes to scattering asymptotic states in a flat background, plane waves of the form
\begin{equation} \label{UsualPlaneWave}
    \psi_{\text{plane}}(t, \vec{r}) = |p_\mp)e^{i\vec{p}\cdot\vec{r}}
\end{equation}
are commonly used to establish boundary conditions at spatial infinity $r\to\infty$ for left-handed and right-handed fermions respectively.
However, under the monopole background \eqref{MonopoleBackground}, the usual plane waves are no longer valid boundary conditions as they do not follow the gauge transformation rule when changing from one patch to another. Explicitly, by considering two coordinate patches with auxiliary vectors $\hat{n}$ and $\hat{n}'$ respectively, the transition of the gauge potential has the form
\begin{equation} \label{LargeGaugePotential}
    \vec{A}_{\hat{n}'}-\vec{A}_{\hat{n}} = -\nabla\zeta, \quad
    \zeta = \arg\left[(n_+|r_+)(r_+|n_+')\right]\in[0,2\pi).
\end{equation}
For wavefunctions on two patches, they are related by
\begin{equation} \label{LargeGaugeWavefunction}
    \psi_{\hat{n}'}(\vec{r}) = e^{i\zeta(\vec{r})}\cdot\psi_{\hat{n}}(\vec{r})
\end{equation}
in the overlap region. Therefore, a valid boundary condition in our case should follow the transformation rule, which the common plane waves \eqref{UsualPlaneWave} do not.

As an alternative, we look for extra factors that generate desired transformations after adding them to the plane waves.
The simplest candidates are $(p_+|r_+)$ and $(p_-|r_+)$ since the transition function of $|r_+)$ in different patches is exactly the same as the one we need for the wavefunction.\footnote{
    See Appendix \ref{SpatialSpinors} and the transformation formula \eqref{LargeGaugeSpinor}.
    The asymmetry between $|r_+)$ and $|r_-)$ originates from the assumption that the product of the electric charge and the magnetic charge is $1/2$ in our system, which implies that the classical electromagnetic angular momentum points from the monopole to the fermion, and therefore $|r_+)$ is preferred.
    If we take the product to be $-1/2$ instead, the direction of the angular momentum would flip, and $|r_-)$ becomes the preferred choice.
} To determine which factor to add, we look at the asymptotic behavior of the function $e^{i\vec{p}\cdot\vec{r}}$ in the large $r$ limit\footnote{The proof is given in Appendix \ref{AsymptoticBehaviorOfPlaneWaves}.}
\begin{equation}
    e^{i\vec{p}\cdot\vec{r}} = \frac{e^{ipr}}{ipr}\delta(1-\hat{p}\cdot\hat{r})+\frac{e^{-ipr}}{-ipr}\delta(1+\hat{p}\cdot\hat{r})+\mathcal{O}\left(\frac{1}{(pr)^2}\right), \quad p = |\vec{p}|.
\end{equation}
At spatial infinity, it is separated into one outgoing part and one incoming part in the leading term.
The outgoing part is localized in the $\hat{p}$ direction, while the incoming part is localized in the $-\hat{p}$ direction.
For incoming fermions, we want to keep the incoming part of the plane wave non-vanishing.
Therefore, the extra factor is $(p_-|r_+)$ since
\begin{equation}
    \left.(p_-|r_+)\right|_{\hat{r}=-\hat{p}} = 1.
\end{equation}
For outgoing fermions, keeping the outgoing part of the plane wave non-vanishing requires the factor to be $(p_+|r_+)$.
Finally, the modified plane waves are
\begin{equation} \label{ModifiedPlaneWaveLIn}
    \psi_{L\vec{p}}^{\text{plane,in}} = |p_-)(p_-|r_+)e^{i\vec{p}\cdot\vec{r}},
\end{equation}
\begin{equation}
    \psi_{L\vec{p}}^{\text{plane,out}} = |p_-)(p_+|r_+)e^{i\vec{p}\cdot\vec{r}},
\end{equation}
\begin{equation}
    \psi_{R\vec{p}}^{\text{plane,in}} = |p_+)(p_-|r_+)e^{i\vec{p}\cdot\vec{r}},
\end{equation}
\begin{equation} \label{ModifiedPlaneWaveROut}
    \psi_{R\vec{p}}^{\text{plane,out}} = |p_+)(p_+|r_+)e^{i\vec{p}\cdot\vec{r}}.
\end{equation}
These are modified asymptotic momentum states that will be considered as in-states and out-states in scattering processes in the following\footnote{
    The extra factor also ensures that the modified plane wave can be expanded on the spin-weighted spherical harmonics properly. For more detailed discussion, see Appendix \ref{AsymptoticBehaviorOfPlaneWaves}.
}. 

Going from wavefunctions to scattering amplitudes, there are two ways to perform calculations.
One is to feed the plane wave into the Dirac equation and solve the evolution perturbatively, as shown in most of the quantum mechanics textbooks.
The other is to find an exact basis for the Dirac equation and utilize it to reconstruct the plane wave in the asymptotic region \cite{Lee_2023}.
In our case, since we already obtain the $(E,\ell,m)$ basis, it is more straightforward to apply the second strategy.
In the first step, we need to determine the expansion coefficients
\begin{equation} \label{PlaneWaveExpansion1}
    \psi_{L\vec{p}}^{\text{in}}(\vec{r}) = \sum_{\ell m}c_{L\ell m}^{\text{in}}\psi_{p\ell m}^{(\hat{k})}(\vec{r}) \xrightarrow{r\to\infty} \psi_{L\vec{p}}^{\text{plane,in}}(\vec{r}),
\end{equation}
\begin{equation} 
    \psi_{L\vec{p}}^{\text{out}}(\vec{r}) = \sum_{\ell m}c_{L\ell m}^{\text{out}}\psi_{p\ell m}^{(\hat{k})}(\vec{r}) \xrightarrow{r\to\infty} \psi_{L\vec{p}}^{\text{plane,out}}(\vec{r}),
\end{equation}
\begin{equation}
    \psi_{R\vec{p}}^{\text{in}}(\vec{r}) = \sum_{\ell m}c_{R\ell m}^{\text{in}}\psi_{(-p)\ell m}^{(\hat{k})}(\vec{r}) \xrightarrow{r\to\infty} \psi_{R\vec{p}}^{\text{plane,in}}(\vec{r}),
\end{equation}
\begin{equation} \label{PlaneWaveExpansion4}
    \psi_{R\vec{p}}^{\text{out}}(\vec{r}) = \sum_{\ell m}c_{R\ell m}^{\text{out}}\psi_{(-p)\ell m}^{(\hat{k})}(\vec{r}) \xrightarrow{r\to\infty} \psi_{R\vec{p}}^{\text{plane,out}}(\vec{r}).
\end{equation}
With the relation \eqref{SpinWeightedIdentity} and the asymptotic formula of the spherical Bessel function 
\begin{equation} \label{BesselAsymptotic}
    j_\nu(Er) \xrightarrow{r\to\infty} \frac{1}{Er}\sin\left(Er-\frac{\pi\nu}{2}\right),
\end{equation}
the coefficients are fixed to be
\begin{equation} \label{InStateExpansionLeft}
     \psi_{L\vec{p}}^{\text{in}}(\vec{r}) = \pi\left[ \sqrt{2} \psi_{p00}^{(-\hat{p})}(\vec{r})+\sum_{\ell=1}^\infty\sqrt{2(2\ell+1)}e^{-\frac{i\pi n}{2}}\psi_{p\ell0}^{(-\hat{p})}(\vec{r})\right],
\end{equation}
\begin{equation} \label{OutStateExpansionLeft}
    \psi_{L\vec{p}}^{\text{out}}(\vec{r}) = \pi\sum_{\ell=1}^\infty\sqrt{2(2\ell+1)}e^{\frac{i\pi n}{2}}\psi_{p\ell(-1)}^{(\hat{p})}(\vec{r}).
\end{equation}
A similar discussion for the right-handed fermion shows
\begin{equation}
    \psi_{R\vec{p}}^{\text{in}}(\vec{r}) = \pi\sum_{\ell=1}^\infty\sqrt{2(2\ell+1)}e^{\frac{-i\pi n}{2}}\psi_{(-p)\ell(-1)}^{(-\hat{p})}(\vec{r}),
\end{equation}
\begin{equation} \label{OutStateExpansionRight}
     \psi_{R\vec{p}}^{\text{out}}(\vec{r}) = \pi\left[ \sqrt{2} \psi_{(-p)00}^{(\hat{p})}(\vec{r})+\sum_{\ell=1}^\infty\sqrt{2(2\ell+1)}e^{\frac{i\pi n}{2}}\psi_{(-p)\ell0}^{(\hat{p})}(\vec{r})\right].
\end{equation}
The relations in \eqref{InStateExpansionLeft}-\eqref{OutStateExpansionRight} connect asymptotic plane waves to the exact $(E,\ell,m)$ basis.

\section{Introduction to Spatial Spinors and Spin-Weighted Spherical Harmonics} \label{SpatialSpinors}

In this appendix, we introduce the spatial spinor and spin-weighted spherical harmonics. We base part of our discussion on the reference \cite{Lee_2023}.

In the partial-wave analysis of the Dirac equation under monopole background, it is convenient to introduce the spatial spinors as the fundamental building blocks of the partial-wave polynomials.
For time-independent wavefunctions, the spatial rotation symmetry group $SO(3)$ has a double cover $SU(2)$, that is, the position vector can be decomposed into 2-component spinors as \cite{Lee_2023}
\begin{equation}
    \hat{r}\cdot\vec{\sigma} = |r_+)(r_+|-|r_-)(r_-|, \quad
    |r_\pm) = [(r_\pm|]^\dagger,
\end{equation}
where $\vec{\sigma}$ are the $2\times2$ Pauli matrices.

To take spatial derivatives of spatial spinors, we need to fix the phase ambiguity $|r_\pm)\to e^{\pm i\phi}|r_\pm)$ first. Here we introduce an auxiliary unit vector $\hat{\eta}\cdot\vec{\sigma} = |\eta_+)(\eta_+|-|\eta_-)(\eta_-|$ as a reference and construct other spinors by applying rotation transformations
\begin{equation} \label{SpatialSpinorDefinition}
    |r_\pm) = U_{\hat{\eta}}(\hat{r})|\eta_\pm) = \exp\left(i\theta\hat{k}\cdot\frac{\vec{\sigma}}{2}\right)|\eta_\pm), \quad
    \theta = \arccos(\hat{r}\cdot\hat{\eta}), \quad \hat{k} = \frac{\hat{\eta}\times\hat{r}}{|\hat{\eta}\times\hat{r}|}
\end{equation}
with $\theta\in[-\pi,\pi)$.
We call $\hat{\eta}$ the reference vector and $|\eta_\pm)$ the reference spinors.
By direct calculations with respect to \eqref{SpatialSpinorDefinition}, some first-order differential equations with the spatial spinor included are as follows:
\begin{equation} \label{SpatialSpinorIdentity1}
    \vec{\sigma}\cdot(-i\nabla)(p|r_\pm) = \mp\frac{i}{r}(p|r_\pm)-\frac{i}{2r}\frac{z_\pm}{1+\hat{r}\cdot\hat{\eta}}(p|r_\mp),
\end{equation}
\begin{equation} \label{SpatialSpinorIdentity3}
    \vec{\sigma}|r_\pm)\cdot(-i\nabla)(p|r_\pm) = -\frac{i}{2r}\frac{z_\pm}{1+\hat{r}\cdot\hat{\eta}}|r_\mp)(p|r_\pm),
\end{equation}
\begin{equation} \label{SpatialSpinorIdentity4}
    \vec{\sigma}|r_\pm)\cdot(-i\nabla)(p|r_\mp) = \frac{i}{2r}\frac{z_\pm}{1+\hat{r}\cdot\hat{\eta}}|r_\mp)(p|r_\pm)\pm\frac{i}{r}|r_\mp)(p|r_\pm),
\end{equation}
with $|p)$ an arbitrary spinor and the factor $z_\pm$ defined as
\begin{equation}
    z_\pm = z_\mp^* = (r_\mp|\hat{\eta}\cdot\vec{\sigma}|r_\pm).
\end{equation}

So far, we discussed properties of spatial spinors under a fixed reference vector.
Now we look at their transformation under a change of reference vector.
By direct calculations, the transformations of the spatial spinors between two reference vectors are
\begin{equation} \label{LargeGaugeSpinor}
    |r_+)_{\hat{\eta}'} = e^{i\bar{\zeta}}|r_+)_{\hat{\eta}} \quad
    |r_-)_{\hat{\eta}'} = e^{-i\bar{\zeta}}|r_-)_{\hat{\eta}}, \quad
    \bar{\zeta} = \arg\left[(\eta_+|r_+)(r_+|\eta_+')\right].
\end{equation}
Note that \eqref{LargeGaugeSpinor} shares many similarities with \eqref{LargeGaugeWavefunction} once we identify the reference vector $\hat{\eta}$ to be the auxiliary vector $\hat{n}$ that indicates the coordinate patch.
For example, we have $\left.\bar{\zeta}\right|_{\hat{\eta}=\hat{n}}=\zeta$ in \eqref{LargeGaugePotential}.
This encourages us to set $\hat{\eta}=\hat{n}$ when discussing Dirac equation and wavefunctions for convenience.

Next, with the help of spatial spinors, we can now introduce the spin-weighted spherical harmonics more easily.
In the definition of the $(E,\ell,m)$ basis, the angular momentum operator acts on fermionic wavefunctions as 
\begin{equation}
    \vec{J}[\psi](\vec{r}) = \left[\vec{r}\times\left(-i\nabla+\vec{A}\right)+\frac{1}{2}\hat{r}+\frac{1}{2}\vec{\sigma}\right]\psi(\vec{r}), \quad
    \vec{A} = \frac{1}{2r}\frac{\vec{r}\times\hat{n}}{r+\vec{r}\cdot\hat{n}}.
\end{equation}
The spinor structure of the wavefunction can be decomposed as
\begin{equation}
    \psi(\vec{r}) = |r_+)\psi_+(\vec{r})+|r_-)\psi_-(\vec{r}).
\end{equation}

Using the identities \eqref{SpatialSpinorIdentity1}-\eqref{SpatialSpinorIdentity4} and identifying the reference vector as the patch direction that appears in the gauge potential, we find that the eigenfunctions of two angular operators $\vec{J}^2$ and $\vec{J}\cdot\hat{k}$ take the forms
\begin{equation} \label{FermionicAngularBasis}
    |r_+)Y_{\ell m}^0(\hat{r}, \hat{k}), \quad |r_-)Y_{\ell m}^1(\hat{r}, \hat{k}),
\end{equation}
where the $Y_{\ell m}^s(\hat{r}, \hat{k})$ are the spin-weighted spherical harmonics defined as:
\begin{equation} \label{SpinWeightedSphericalHarmonics}
    Y_{\ell m}^s(\hat{r}, \hat{k}) =  \mathcal{C}_{\ell m}^s\big[(k_-|^{(i_1}\cdots(k_-|^{i_{\ell+m}}(k_+|^{i_{\ell+m+1}}\cdots(k_+|^{i_{2\ell})}\big]\big[|r_+)_{(i_1}\cdots|r_+)_{i_{\ell + s}}|r_-)_{i_{\ell +s+1}}\cdots|r_-)_{i_{2\ell})}\big],
\end{equation}
where
\begin{equation}
    C_{\ell m}^s = \left[\frac{1}{4\pi}\frac{(2\ell )!(2\ell+1)!}{(\ell+m)!(\ell-m)!(\ell+s)!(\ell-s)!}\right]^{1/2}.
\end{equation}
The indices run as $m,s = -\ell, -\ell + 1, \ldots,\ell$, and the parenthesis over indices stand for total symmetrization with the symmetric factor divided.

The normalization of the spin-weighted spherical harmonics is
\begin{equation}
    \int d^2\hat{r} \ (Y_{jm}^s)^*(\hat{r}, \hat{k})Y_{j'm'}^s(\hat{r}, \hat{k}) = \delta_{jj'}\delta_{mm'}.
\end{equation}
Note that comparing to the $(E,\ell,m)$ basis given in Appendix \ref{ModeExpansionUnderMonopoleBackground}, two angular functions with the same quantum numbers in \eqref{FermionicAngularBasis} are mixed with different radial functions to solve the Dirac equation as in \eqref{PositiveEnergyMode1}.
The identity relation is given by
\begin{equation} \label{SpinWeightedIdentity}
    \delta(\hat{r}\cdot\hat{k}-1) = \sum_{j=|s|}^{\infty}\sqrt{\pi(2j+1)}Y_{j(-s)}^s(\hat{r}, \hat{k}),
\end{equation}
which is important when calculating the plane wave expansion in \eqref{PlaneWaveExpansion1}-\eqref{PlaneWaveExpansion4}.

So far, the discussion for spin-weighted spherical harmonics is for fixed $\hat{k}$. 
Switching to a different $\hat{k}$ in the $(E,\ell,m)$ basis requires us to find a new set of spherical functions.
For example, taking $\hat{k}_1=\hat{z}$ and $\hat{k}_2=\hat{x}$ corresponds to labeling functions by $(J^2,J_z)$ and $(J^2,J_x)$ respectively, which gives different eigenfunctions that are linearly related.
Following the definition in \eqref{SpinWeightedSphericalHarmonics}, we have
\begin{equation} \label{INTEGRALTWOYS}
    \int d^2\hat{r} \ (Y_{jm}^s)^*(\hat{r}, \hat{k})Y_{j'm'}^s(\hat{r}, \hat{k}') = \delta_{jj'}\cdot\sqrt{\frac{4\pi}{2j+1}}Y_{j m'}^{-m}(\hat{k},\hat{k}').
\end{equation}

This provides the linear transformation between bases of different $\hat{k}$:
\begin{equation} \label{SpinWeightedKChanged}
    Y_{jm}^s(\hat{r},\hat{k}) = \sqrt{\frac{4\pi}{2j+1}}\sum_{m'}Y_{j m}^{-m'}(\hat{k}',\hat{k})Y_{jm'}^s(\hat{r},\hat{k}').
\end{equation}

\section{Asymptotic Behavior of Plane Waves} \label{AsymptoticBehaviorOfPlaneWaves}

When the system is free, we use Legendre polynomials $P_\nu(x)$ to perform partial-wave expansions.
We have
\begin{equation}
    \psi_{\vec{p}}^{\text{plane}}(\vec{r}) = e^{i\vec{p}\cdot\vec{r}} = \sum_{\ell=0}^\infty (2\ell+1)i^\ell j_\ell(pr)P_\ell(\hat{p}\cdot\hat{r}).
\end{equation}
With \eqref{BesselAsymptotic}, we can rewrite the plane wave in the asymptotic region as
\begin{equation} \label{AsymptoticPlaneWave}
    \begin{split}
        e^{i\vec{p}\cdot\vec{r}}
        &= \sum_{\ell=0}^\infty (2\ell+1)i^\ell \left[\frac{\exp\left(ipr-\frac{i\pi\ell}{2}\right)-\exp\left(-ipr+\frac{i\pi\ell}{2}\right)}{2ipr}+\mathcal{O}\left(\frac{1}{(pr)^2}\right)\right]P_\ell(\hat{p}\cdot\hat{r}) \\
        &= \frac{e^{ipr}}{ipr}\sum_{\ell=0}^\infty \frac{2\ell+1}{2}P_\ell(\hat{p}\cdot\hat{r})+\frac{e^{-ipr}}{-ipr}\sum_{\ell=0}^\infty \frac{2\ell+1}{2}(-1)^\ell P_\ell(\hat{p}\cdot\hat{r})+\mathcal{O}\left(\frac{1}{(pr)^2}\right) \\
        &= \frac{e^{ipr}}{ipr}\delta(1-\hat{p}\cdot\hat{r})+\frac{e^{-ipr}}{-ipr}\delta(1+\hat{p}\cdot\hat{r})+\mathcal{O}\left(\frac{1}{(pr)^2}\right).
    \end{split}
\end{equation}

When there is a monopole in the background, we no longer choose Legendre polynomials as our partial-wave basis.
Instead, we use spin-weighted spherical harmonics as given in \eqref{FermionicAngularBasis}. 
For later convenience, we take $\hat{k}=\hat{p}$ and project the spinor structure onto the basis $|p_\pm)$. 
Then \eqref{FermionicAngularBasis} can be decomposed into
\begin{equation}
    |p_+)Y_{\ell m}^{1/2}(\hat{r}, \hat{p}), \quad
    |p_-)Y_{\ell m}^{1/2}(\hat{r}, \hat{p})
\end{equation}
with $\ell$ and $m$ being half-integers.
In the following discussion, we consider only $m=1/2$  since these spin-weighted spherical harmonics with other values of $m$ have extra dependence on the azimuthal angle $\phi$,\footnote{
    Taking $\hat{p}=\hat{z}$ we parametrize the position vector using the spherical coordinates as
    \begin{equation}
        \vec{r}=(r\sin\theta\cos\phi, r\sin\theta\sin\phi, r\cos\theta),
    \end{equation}
which defines the azimuthal angle.} which does not appear in the plane wave.
They are called the monopole harmonics \cite{Wu:1976ge}
\begin{equation}
    \mathcal{P}_\nu(\hat{r}, \hat{p}) = \sqrt{4\pi}Y_{(\nu+1/2)1/2}^{1/2}(\hat{r}, \hat{p}).
\end{equation}

The monopole harmonics have several different properties compared to the Legendre polynomials $P_\nu(\hat{p}\cdot\hat{r})=\sqrt{4\pi}Y_{\nu0}^0(\hat{r},\hat{p})$.
First, it is not invariant under the transformation of the reference vector in \eqref{LargeGaugeSpinor}.
It can be separated into a covariant part and an invariant part as
\begin{equation}
    \mathcal{P}_\nu(\hat{r}, \hat{p}) = (p_+|r_+)\cdot\tilde{\mathcal{P}}_\nu(\hat{p}\cdot\hat{r}),
\end{equation}
where $\tilde{\mathcal{P}}_\nu$ is a single-variable function.
This covariant factor ensures that the partial-wave basis transforms properly under different coordinate patches.
Second, its infinite sums are
\begin{equation}
    \begin{split}
        \sum_{\nu=0}^\infty \frac{2(\nu+1/2)+1}{2}\mathcal{P}_\nu(\hat{p}\cdot\hat{r}) &= (p_+|r_+)\sum_{\nu=0}^\infty \frac{2(\nu+1/2)+1}{2}\tilde{\mathcal{P}}_\nu(\hat{p}\cdot\hat{r}) \\ 
        &= (p_+|r_+)\cdot\delta(1-\hat{p}\cdot\hat{r}) = \delta(1-\hat{p}\cdot\hat{r}),
    \end{split}
\end{equation}
\begin{equation}
    \begin{split}
        \sum_{\nu=0}^\infty \frac{2(\nu+1/2)+1}{2}(-1)^{\nu}\mathcal{P}_\nu(\hat{p}\cdot\hat{r}) &= (p_+|r_+)\sum_{\nu=0}^\infty \frac{2(\nu+1/2)+1}{2}(-1)^{\nu}\tilde{\mathcal{P}}_\nu(\hat{p}\cdot\hat{r}) \\ 
        &= (p_+|r_+)\cdot\delta(1+\hat{p}\cdot\hat{r}) = 0.
    \end{split}
\end{equation}
Because of the extra factor, we cannot reconstruct two delta functions in opposite directions as in \eqref{AsymptoticPlaneWave} using the monopole harmonics.
Therefore, to match the angular delta functions, we require the plane wave to be modified as
\begin{equation}
    \psi_{\vec{p}}^{\text{modified plane}}(\vec{r}) = (p_+|r_+)e^{i\vec{p}\cdot\vec{r}}.
\end{equation}
This ensures that the modified plane wave can be expanded on the monopole harmonics. Let us also observe that, as indicated in Section \ref{Section4.1}, this additional factor also endows the solution with the appropriate behavior through patch transitions.

\section{A Twisted Ising Example} \label{IsingExample}

In this appendix we provide a pedagogical example of the formalism of twisted boundary states in the context of the Ising model. We notice that the results summarized in this appendix were obtained first in \cite{Watts:2000kp} using the Kramers-Wannier duality of the Ising model, rather than the formalism of twisted Cardy states.

The boundary states for the Ising model are:
\begin{align}
    |0\rangle &= \frac{1}{\sqrt{2}} |0\rrangle + \frac{1}{\sqrt{2}} | \epsilon \rrangle + \frac{1}{2^{1/4}} | \sigma \rrangle, \label{Ising0boundarystate} \\[0.3cm]
    | \epsilon \rangle &= \frac{1}{\sqrt{2}} |0\rrangle + \frac{1}{\sqrt{2}} | \epsilon \rrangle - \frac{1}{2^{1/4}} | \sigma \rrangle, \label{Isingepsilonboundarystate} \\[0.3cm]
    | \sigma \rangle &= |0\rrangle - | \epsilon \rrangle. \label{Isingsigmaboundarystate}
\end{align}
On top of these, the $\sigma$-boundary allows for a $\mathbb{Z}_{2}$-twisted sector. This is, the $\mathbb{Z}_{2}$ line $\eta$  can end topologically at the $\sigma$ boundary. Concretely, we have
\begin{equation} \label{Z2Twistedsectorsigmaboundary}
    | \sigma \rangle_{\eta} = \alpha | \mu \rrangle,
\end{equation}
where we have only one diagonal element in the $\mathbb{Z}_{2}$ twisted sector. To see this, recall that the $\mathbb{Z}_{2}$ twisted sector contains three operators: the holomorphic fermion $\psi(z)$, the antiholomorphic fermion $\bar{\psi}(\bar{z})$, and the disorder operator $\mu(z, \bar{z})$, of which only $\mu(z, \bar{z})$ is diagonal (and thus can contribute to an Ishibashi). Explicitly, this can be seen by calculating the partition function in the $\mathbb{Z}_{2}$ twisted sector:
\begin{equation}
    Z[01] = \bar{\chi}_{0} \chi_{\frac{1}{2}} + \bar{\chi}_{\frac{1}{2}} \chi_{0} + \bar{\chi}_{\frac{1}{16}} \chi_{\frac{1}{16}}.
\end{equation}
Clearly, only one primary is diagonal in between the holomorphic and antiholomorphic sides, and it corresponds to the disorder operator $\mu(z, \bar{z})$.

We can determine the normalization constant $\alpha$ using the generalized Cardy condition (recall Eqn. \eqref{twistedmodularity}):
\begin{equation}
    \langle \sigma_{\eta} | \tilde{q}^{\frac{1}{2}(L_{0} + \bar{L}_{0} - c/12) } | \sigma_{\eta} \rangle = \alpha^{2} \chi_{\sigma}(\tilde{q}) = \frac{\alpha^{2}}{\sqrt{2}}( \chi_{0}(q) - \chi_{\epsilon}(q) ).
\end{equation}
Thus, to get integers on the right-hand side we need $\alpha = 2^{1/4}$:
\begin{equation}
    | \sigma \rangle_{\eta} = 2^{1/4} | \mu \rrangle.
\end{equation}

One-point functions of local operators in presence of a boundary can be calculated from \eqref{onepointcircleoverlap}. We can proceed similarly for an $a$-symmetric boundary condition $B$ with twisted boundary sector $|B\rangle_{a}$ and quasi-local operators $\varphi_{a}$ living on a $a$-twisted sector:
\begin{equation}
    \langle \varphi_{a}(z, \bar{z}) \rangle_{B} = \frac{\langle B_{a} | \varphi_{a} \rangle/ \langle B | 0 \rangle}{|z - \bar{z}|^{2h_{\varphi_{a}}}}.
\end{equation}
Then, we have the one-point functions in the twisted sector:
\begin{equation} \label{result}
    \langle \psi(z) \rangle_{\sigma} = \langle \bar{\psi}(\bar{z}) \rangle_{\sigma} = 0, \quad \langle \mu(z,\bar{z}) \rangle_{\sigma} = \frac{2^{1/4}}{|z - \bar{z}|^{1/8}}.
\end{equation}
In particular, notice that the one-point function coefficient of $\langle \mu(z, \bar{z}) \rangle_{\sigma}$ cannot be recovered from the standard boundary states in \eqref{Ising0boundarystate}-\eqref{Isingsigmaboundarystate}, and it follows from the particular normalization of the $\sigma$ boundary condition on the twisted sector: $|\sigma \rangle_{\eta}$.

\bibliographystyle{JHEP}
\bibliography{references}

@article{Nussinov:1982BaryonAsymmetry,
    author = "Nussinov, S.",
    title = "{Superheavy Monopoles and Baryon Asymmetry}",
    journal = "Phys. Lett. B",
    volume = "110",
    pages = "221--226",
    year = "1982"
}

@article{Ellis:1982CatalysisBAU,
    author = "Ellis, John R. and Nanopoulos, Dimitri V. and Olive, Keith A.",
    title = "{Baryon Number Violation Catalyzed by Grand Unified Monopoles}",
    doi = "10.1016/0370-2693(82)90990-X",
    journal = "Phys. Lett. B",
    volume = "116",
    pages = "127--132",
    year = "1982"
}

@article{LangackerPi:1980,
    author = "Langacker, Paul and Pi, So-Young",
    title = "{Magnetic Monopoles in Grand Unified Theories}",
    doi = "10.1103/PhysRevLett.45.1",
    journal = "Phys. Rev. Lett.",
    volume = "45",
    pages = "1--4",
    year = "1980"
}

@article{Davis:1992MonopoleBaryogenesis,
    author = "Davis, Anne-Christine and Earnshaw, Michael A. and Wiedmann, Urs A.",
    title = "{Monopole Baryogenesis in the Langacker-Pi Scenario}",
    doi = "10.1016/0370-2693(92)91489-V",
    journal = "Phys. Lett. B",
    volume = "293",
    pages = "123--126",
    year = "1992"
}

@article{Brennan:2024MonopoleBaryogenesis,
    author = "Brennan, T. Daniel and Wang, Lian-Tao and Xiao, Huangyu",
    title = "{Monopole Catalyzed Baryogenesis with a $\theta$ angle}",
    eprint = "2412.14239",
    archivePrefix = "arXiv",
    primaryClass = "hep-ph",
    year = "2024"
}

@article{tHooft:1974Monopole,
    author = "'t Hooft, Gerard",
    title = "{Magnetic Monopoles in Unified Gauge Theories}",
    doi = "10.1016/0550-3213(74)90486-6",
    journal = "Nucl. Phys. B",
    volume = "79",
    pages = "276--284",
    year = "1974"
}

@article{Polyakov:1974Monopole,
    author = "Polyakov, Alexander M.",
    title = "{Particle Spectrum in Quantum Field Theory}",
    journal = "JETP Lett.",
    volume = "20",
    pages = "194--195",
    year = "1974",
    note = "[Pisma Zh. Eksp. Teor. Fiz. 20 (1974) 430--433]"
}

@article{GoddardOlive:1978,
    author = "Goddard, Peter and Olive, David I.",
    title = "{New Developments in the Theory of Magnetic Monopoles}",
    doi = "10.1088/0034-4885/41/9/001",
    journal = "Rept. Prog. Phys.",
    volume = "41",
    pages = "1357",
    year = "1978"
}

@article{Preskill:1979Monopoles,
    author = "Preskill, John",
    title = "{Cosmological Production of Superheavy Magnetic Monopoles}",
    doi = "10.1103/PhysRevLett.43.1365",
    journal = "Phys. Rev. Lett.",
    volume = "43",
    pages = "1365--1368",
    year = "1979"
}

@article{Preskill:1984Monopoles,
    author = "Preskill, John",
    title = "{Magnetic Monopoles}",
    doi = "10.1146/annurev.ns.34.120184.002333",
    journal = "Ann. Rev. Nucl. Part. Sci.",
    volume = "34",
    pages = "461--530",
    year = "1984"
}

@article{Rubakov:1981Monopole,
    author = "Rubakov, V. A.",
    title = "{Superheavy Magnetic Monopoles and Proton Decay}",
    journal = "JETP Lett.",
    volume = "33",
    pages = "644--646",
    year = "1981",
    note = "[Pisma Zh. Eksp. Teor. Fiz. 33 (1981) 658--660]"
}

@article{Callan:1983Catalysis,
    author = "Callan, Jr., Curtis G.",
    title = "{Monopole Catalysis of Baryon Decay}",
    doi = "10.1016/0550-3213(83)90677-6",
    journal = "Nucl. Phys. B",
    volume = "212",
    pages = "391--400",
    year = "1983"
}

@article{RubakovSerebryakov:1983,
    author = "Rubakov, V. A. and Serebryakov, M. S.",
    title = "{Anomalous Baryon Number Nonconservation in the Presence of SU(5) Monopoles}",
    doi = "10.1016/0550-3213(83)90484-4",
    journal = "Nucl. Phys. B",
    volume = "218",
    pages = "240--268",
    year = "1983"
}

@article{DawsonSchellekens:1983,
    author = "Dawson, Sally and Schellekens, A. N.",
    title = "{Monopole Catalysis of Proton Decay in SO(10) Grand Unified Models}",
    doi = "10.1103/PhysRevD.27.2119",
    journal = "Phys. Rev. D",
    volume = "27",
    pages = "2119",
    year = "1983"
}

@article{DimopoulosPreskillWilczek:1982,
    author = "Dimopoulos, Savas and Preskill, John and Wilczek, Frank",
    title = "{Catalyzed Nucleon Decay in Neutron Stars}",
    doi = "10.1016/0370-2693(82)90679-7",
    journal = "Phys. Lett. B",
    volume = "119",
    pages = "320--322",
    year = "1982"
}

@article{KolbColgateHarvey:1982,
    author = "Kolb, Edward W. and Colgate, Stirling A. and Harvey, Jeffrey A.",
    title = "{Monopole Catalysis of Nucleon Decay in Neutron Stars}",
    doi = "10.1103/PhysRevLett.49.1373",
    journal = "Phys. Rev. Lett.",
    volume = "49",
    pages = "1373",
    year = "1982"
}

@article{Soudan:1987Monopole,
    author = "Bartelt, J. and Courant, H. and Heller, K. and Joyce, T. and Marshak, M. and Peterson, E. and Ruddick, K. and Shupe, M. and Ayres, D. S. and Dawson, J. W. and Fields, T. H. and May, E. N. and Price, L. E.",
    title = "{Monopole-flux and Proton-decay Limits from the Soudan 1 Detector}",
    doi = "10.1103/PhysRevD.36.1990",
    journal = "Phys. Rev. D",
    volume = "36",
    pages = "1990--2000",
    year = "1987"
}

@article{IMB:1994Monopole,
    author = "Becker-Szendy, R. and others",
    title = "{New Magnetic Monopole Flux Limits from the IMB Proton Decay Detector}",
    doi = "10.1103/PhysRevD.49.2169",
    journal = "Phys. Rev. D",
    volume = "49",
    pages = "2169--2173",
    year = "1994"
}

@article{MACRO:2002Catalysis,
    author = "Ambrosio, M. and others",
    collaboration = "MACRO",
    title = "{Search for Nucleon Decays Induced by GUT Magnetic Monopoles with the MACRO Experiment}",
    eprint = "hep-ex/0207024",
    archivePrefix = "arXiv",
    doi = "10.1140/epjc/s2002-01045-x",
    journal = "Eur. Phys. J. C",
    volume = "26",
    pages = "163--172",
    year = "2002"
}

@article{SuperK:2012Monopole,
    author = "Ueno, K. and others",
    collaboration = "Super-Kamiokande",
    title = "{Search for GUT Monopoles at Super-Kamiokande}",
    eprint = "1203.0940",
    archivePrefix = "arXiv",
    primaryClass = "hep-ex",
    doi = "10.1016/j.astropartphys.2012.05.008",
    journal = "Astropart. Phys.",
    volume = "36",
    pages = "131--136",
    year = "2012"
}

@article{IceCube:2014Monopole,
    author = "Aartsen, M. G. and others",
    collaboration = "IceCube",
    title = "{Search for Non-Relativistic Magnetic Monopoles with IceCube}",
    eprint = "1402.3460",
    archivePrefix = "arXiv",
    primaryClass = "astro-ph.HE",
    doi = "10.1140/epjc/s10052-014-2938-8",
    journal = "Eur. Phys. J. C",
    volume = "74",
    number = "7",
    pages = "2938",
    year = "2014"
}

@article{Candela:2025Monopoles,
    author = "Candela, Pablo M. and Khoze, Valentin V. and Turner, Jessica",
    title = "{Monopoles at Future Neutrino Detectors}",
    eprint = "2504.14918",
    archivePrefix = "arXiv",
    primaryClass = "hep-ph",
    doi = "10.1007/JHEP07(2025)034",
    journal = "JHEP",
    volume = "07",
    pages = "034",
    year = "2025"
}

@Article{Ishibashi:1988kg,
  author        = {Ishibashi, Nobuyuki},
  title         = {The Boundary and Crosscap States in Conformal Field Theories},
  journal       = {Mod. Phys. Lett. A},
  doi = {10.1142/S0217732389000320},
  url  = {https://doi.org/10.1142/S0217732389000320},
  volume = "4",
  pages = "251",
  year = "1989"
}

@book{recknagel_schomerus_2013, place={Cambridge}, series={Cambridge Monographs on Mathematical Physics}, title={Boundary Conformal Field Theory and the Worldsheet Approach to D-Branes}, DOI={10.1017/CBO9780511806476}, publisher={Cambridge University Press}, author={Recknagel, Andreas and Schomerus, Volker}, year={2013}, collection={Cambridge Monographs on Mathematical Physics}}

@article{Fukusumi:2021zme,
    author = "Fukusumi, Yoshiki and Tachikawa, Yuji and Zheng, Yunqin",
    title = "{Fermionization and boundary states in 1+1 dimensions}",
    eprint = "2103.00746",
    archivePrefix = "arXiv",
    primaryClass = "hep-th",
    doi = "10.21468/SciPostPhys.11.4.082",
    journal = "SciPost Phys.",
    volume = "11",
    number = "4",
    pages = "082",
    year = "2021"
}

@article{Ebisu:2021acm,
    author = "Ebisu, Hiromi and Watanabe, Masataka",
    title = "{Fermionization of conformal boundary states}",
    eprint = "2103.01101",
    archivePrefix = "arXiv",
    primaryClass = "hep-th",
    doi = "10.1103/PhysRevB.104.195124",
    journal = "Phys. Rev. B",
    volume = "104",
    number = "19",
    pages = "195124",
    year = "2021"
}

@book{DiFrancesco:1997nk,
    author = "Di Francesco, P. and Mathieu, P. and Senechal, D.",
    title = "{Conformal Field Theory}",
    doi = "10.1007/978-1-4612-2256-9",
    isbn = "978-0-387-94785-3, 978-1-4612-7475-9",
    publisher = "Springer-Verlag",
    address = "New York",
    series = "Graduate Texts in Contemporary Physics",
    year = "1997"
}

@article{CARDY1991274,
title = {Bulk and boundary operators in conformal field theory},
journal = {Physics Letters B},
volume = {259},
number = {3},
pages = {274-278},
year = {1991},
issn = {0370-2693},
doi = {https://doi.org/10.1016/0370-2693(91)90828-E},
url = {https://www.sciencedirect.com/science/article/pii/037026939190828E},
author = {John L. Cardy and David C. Lewellen}
}

@article{Watts:2000kp,
    author = "Watts, G. M. T.",
    title = "{On the boundary Ising model with disorder operators}",
    eprint = "hep-th/0002218",
    archivePrefix = "arXiv",
    reportNumber = "KCL-MTH-00-06",
    doi = "10.1016/S0550-3213(00)00720-3",
    journal = "Nucl. Phys. B",
    volume = "596",
    pages = "513--524",
    year = "2001"
}

@article{Kusuki:2023bsp,
    author = "Kusuki, Yuya and Murciano, Sara and Ooguri, Hirosi and Pal, Sridip",
    title = "{Symmetry-resolved entanglement entropy, spectra \& boundary conformal field theory}",
    eprint = "2309.03287",
    archivePrefix = "arXiv",
    primaryClass = "hep-th",
    reportNumber = "CALT-TH 2023-037, IPMU 23-0030, RIKEN-iTHEMS-Report-23",
    doi = "10.1007/JHEP11(2023)216",
    journal = "JHEP",
    volume = "11",
    pages = "216",
    year = "2023"
}

@article{vanBeest:2023dbu,
    author = "van Beest, Marieke and Boyle Smith, Philip and Delmastro, Diego and Komargodski, Zohar and Tong, David",
    title = "{Monopoles, Scattering, and Generalized Symmetries}",
    eprint = "2306.07318",
    archivePrefix = "arXiv",
    primaryClass = "hep-th",
    month = "6",
    year = "2023"
}

@article{vanBeest:2023mbs,
    author = "van Beest, Marieke and Boyle Smith, Philip and Delmastro, Diego and Mouland, Rishi and Tong, David",
    title = "{Fermion-monopole scattering in the Standard Model}",
    eprint = "2312.17746",
    archivePrefix = "arXiv",
    primaryClass = "hep-th",
    doi = "10.1007/JHEP08(2024)004",
    journal = "JHEP",
    volume = "08",
    pages = "004",
    year = "2024"
}

@article{Lee_2023,
   title={Classical observables from partial wave amplitudes},
   volume={2023},
   ISSN={1029-8479},
   url={http://dx.doi.org/10.1007/JHEP06(2023)096},
   DOI={10.1007/jhep06(2023)096},
   number={6},
   journal={Journal of High Energy Physics},
   publisher={Springer Science and Business Media LLC},
   author={Lee, Hojin and Lee, Sangmin and Mazumdar, Subhajit},
   year={2023},
   month=jun }

@article{PhysRevD.15.2287,
  title = {Scattering of a Dirac particle with charge $Ze$ by a fixed magnetic monopole},
  author = {Kazama, Yoichi and Yang, Chen Ning and Goldhaber, Alfred S.},
  journal = {Phys. Rev. D},
  volume = {15},
  issue = {8},
  pages = {2287--2299},
  numpages = {0},
  year = {1977},
  month = {Apr},
  publisher = {American Physical Society},
  doi = {10.1103/PhysRevD.15.2287},
  url = {https://link.aps.org/doi/10.1103/PhysRevD.15.2287}
}

@article{Wu:1976ge,
    author = "Wu, Tai Tsun and Yang, Chen Ning",
    title = "{Dirac Monopole Without Strings: Monopole Harmonics}",
    reportNumber = "ITP-SB-76-5",
    doi = "10.1016/0550-3213(76)90143-7",
    journal = "Nucl. Phys. B",
    volume = "107",
    pages = "365",
    year = "1976"
}

@article{Cardy:1989ir,
    author = "Cardy, John L.",
    title = "{Boundary Conditions, Fusion Rules and the Verlinde Formula}",
    reportNumber = "UCSB-TH-89-06",
    doi = "10.1016/0550-3213(89)90521-X",
    journal = "Nucl. Phys. B",
    volume = "324",
    pages = "581--596",
    year = "1989"
}

@article{Callan:1982ah,
    author = "Callan, Jr., Curtis G.",
    title = "{Disappearing Dyons}",
    reportNumber = "Print-82-0006 (PRINCETON)",
    doi = "10.1103/PhysRevD.25.2141",
    journal = "Phys. Rev. D",
    volume = "25",
    pages = "2141",
    year = "1982"
}

@article{Callan:1982au,
    author = "Callan, Jr., Curtis G.",
    title = "{Dyon-Fermion Dynamics}",
    reportNumber = "Print-82-0476 (PRINCETON)",
    doi = "10.1103/PhysRevD.26.2058",
    journal = "Phys. Rev. D",
    volume = "26",
    pages = "2058--2068",
    year = "1982"
}

@article{Callan:1983ed,
    author = "Callan, Jr., Curtis G. and Das, Sumit R.",
    title = "{Boundary Conditions on the Monopole Dirac Equation}",
    reportNumber = "EFI-83-36-CHICAGO",
    doi = "10.1103/PhysRevLett.51.1155",
    journal = "Phys. Rev. Lett.",
    volume = "51",
    pages = "1155",
    year = "1983"
}

@article{Rubakov:1982fp,
    author = "Rubakov, V. A.",
    title = "{Adler-Bell-Jackiw Anomaly and Fermion Number Breaking in the Presence of a Magnetic Monopole}",
    doi = "10.1016/0550-3213(82)90034-7",
    journal = "Nucl. Phys. B",
    volume = "203",
    pages = "311--348",
    year = "1982"
}

@article{Rubakov:1988aq,
    author = "Rubakov, V. A.",
    title = "{Monopole Catalysis of Proton Decay}",
    doi = "10.1088/0034-4885/51/2/002",
    journal = "Rept. Prog. Phys.",
    volume = "51",
    pages = "189--241",
    year = "1988"
}

@article{Polchinski:1984uw,
    author = "Polchinski, Joseph",
    title = "{Monopole Catalysis: The Fermion Rotor System}",
    reportNumber = "HUTP-84-A005",
    doi = "10.1016/0550-3213(84)90398-5",
    journal = "Nucl. Phys. B",
    volume = "242",
    pages = "345--363",
    year = "1984"
}

@article{Choi:2024tri,
    author = "Choi, Yichul and Rayhaun, Brandon C. and Zheng, Yunqin",
    title = "{Generalized Tube Algebras, Symmetry-Resolved Partition Functions, and Twisted Boundary States}",
    eprint = "2409.02159",
    archivePrefix = "arXiv",
    primaryClass = "hep-th",
    month = "9",
    year = "2024"
}

@article{Choi:2024wfm,
    author = "Choi, Yichul and Rayhaun, Brandon C. and Zheng, Yunqin",
    title = "{Noninvertible Symmetry-Resolved Affleck-Ludwig-Cardy Formula and Entanglement Entropy from the Boundary Tube Algebra}",
    eprint = "2409.02806",
    archivePrefix = "arXiv",
    primaryClass = "hep-th",
    doi = "10.1103/PhysRevLett.133.251602",
    journal = "Phys. Rev. Lett.",
    volume = "133",
    number = "25",
    pages = "251602",
    year = "2024"
}

@article{Choi:2023xjw,
    author = "Choi, Yichul and Rayhaun, Brandon C. and Sanghavi, Yaman and Shao, Shu-Heng",
    title = "{Remarks on boundaries, anomalies, and noninvertible symmetries}",
    eprint = "2305.09713",
    archivePrefix = "arXiv",
    primaryClass = "hep-th",
    doi = "10.1103/PhysRevD.108.125005",
    journal = "Phys. Rev. D",
    volume = "108",
    number = "12",
    pages = "125005",
    year = "2023"
}

@article{Collier:2021ngi,
    author = "Collier, Scott and Mazac, Dalimil and Wang, Yifan",
    title = "{Bootstrapping boundaries and branes}",
    eprint = "2112.00750",
    archivePrefix = "arXiv",
    primaryClass = "hep-th",
    doi = "10.1007/JHEP02(2023)019",
    journal = "JHEP",
    volume = "02",
    pages = "019",
    year = "2023"
}

@article{Smith:2019jnh,
    author = "Smith, Philip Boyle and Tong, David",
    title = "{Boundary States for Chiral Symmetries in Two Dimensions}",
    eprint = "1912.01602",
    archivePrefix = "arXiv",
    primaryClass = "hep-th",
    doi = "10.1007/JHEP09(2020)018",
    journal = "JHEP",
    volume = "09",
    pages = "018",
    year = "2020"
}

@article{Loladze:2025jsq,
    author = "Loladze, Vazha and Okui, Takemichi and Tong, David",
    title = "{Dynamics of the Fermion-Rotor System}",
    eprint = "2508.21059",
    archivePrefix = "arXiv",
    primaryClass = "hep-th",
    month = "8",
    year = "2025"
}

@article{Hamada:2022eiv,
    author = "Hamada, Yuta and Kitahara, Teppei and Sato, Yoshiki",
    title = "{Monopole-fermion scattering and varying Fock space}",
    eprint = "2208.01052",
    archivePrefix = "arXiv",
    primaryClass = "hep-th",
    reportNumber = "KEK-TH-2423, YITP-22-44, TU-1156",
    doi = "10.1007/JHEP11(2022)116",
    journal = "JHEP",
    volume = "11",
    pages = "116",
    year = "2022"
}

@article{Ueda:2025ecm,
    author = "Ueda, Atsushi and Linden, Vic Vander and Lootens, Laurens and Haegeman, Jutho and Fendley, Paul and Verstraete, Frank",
    title = "{Perfect Particle Transmission through Duality Defects}",
    eprint = "2510.26780",
    archivePrefix = "arXiv",
    primaryClass = "hep-th",
    month = "10",
    year = "2025"
}

@article{Tachikawa:2026cxd,
    author = "Tachikawa, Yuji and Tsuji, Keita and Watanabe, Masataka",
    title = "{What happens to wavepackets of fermions when scattered by the Maldacena-Ludwig wall?}",
    eprint = "2603.25508",
    archivePrefix = "arXiv",
    primaryClass = "hep-th",
    month = "3",
    year = "2026"
}

@article{Affleck:1993np,
    author = "Affleck, Ian and Sagi, Jacob",
    title = "{Monopole catalyzed baryon decay: A Boundary conformal field theory approach}",
    eprint = "hep-th/9311056",
    archivePrefix = "arXiv",
    reportNumber = "UBCTP-93-18",
    doi = "10.1016/0550-3213(94)90478-2",
    journal = "Nucl. Phys. B",
    volume = "417",
    pages = "374--402",
    year = "1994"
}

@article{Maldacena:1995pq,
    author = "Maldacena, Juan Martin and Ludwig, A. W. W.",
    title = "{Majorana fermions, exact mapping between quantum impurity fixed points with four bulk fermion species, and solution of the 'unitarity puzzle'}",
    eprint = "cond-mat/9502109",
    archivePrefix = "arXiv",
    doi = "10.1016/S0550-3213(97)00596-8",
    journal = "Nucl. Phys. B",
    volume = "506",
    pages = "565--588",
    year = "1997"
}

@article{Wei:2026fsn,
    author = "Wei, Pengcheng and Zheng, Yunqin",
    title = "{Non-invertible Symmetries in Weyl Fermions, and Applications to Fermion-Boundary Scattering Problem}",
    eprint = "2605.19363",
    archivePrefix = "arXiv",
    primaryClass = "hep-th",
    month = "5",
    year = "2026"
}

@article{Arias-Tamargo:2026urw,
    author = "Arias-Tamargo, Guillermo and Boyle Smith, Philip and Mouland, Rishi and Vel{\'a}squez Cotini Hutt, Maxwell L.",
    title = "{Non-Invertible Symmetries and Boundaries for Two-Dimensional Fermions}",
    eprint = "2605.13952",
    archivePrefix = "arXiv",
    primaryClass = "hep-th",
    month = "5",
    year = "2026"
}

@article{Antinucci:2026uuh,
    author = "Antinucci, Andrea and Copetti, Christian and Galati, Giovanni and Rizi, Giovanni",
    title = "{A Twist on Scattering from Defect Anomalies}",
    eprint = "2605.13961",
    archivePrefix = "arXiv",
    primaryClass = "hep-th",
    month = "5",
    year = "2026"
}

@article{Gaiotto:2014kfa,
    author = "Gaiotto, Davide and Kapustin, Anton and Seiberg, Nathan and Willett, Brian",
    title = "{Generalized Global Symmetries}",
    eprint = "1412.5725",
    archivePrefix = "arXiv",
    primaryClass = "hep-th",
    doi = "10.1007/JHEP02(2015)172",
    journal = "JHEP",
    volume = "02",
    pages = "172",
    year = "2015"
}

@article{Chang:2018iay,
    author = "Chang, Chi-Ming and Lin, Ying-Hsuan and Shao, Shu-Heng and Wang, Yifan and Yin, Xi",
    title = "{Topological Defect Lines and Renormalization Group Flows in Two Dimensions}",
    eprint = "1802.04445",
    archivePrefix = "arXiv",
    primaryClass = "hep-th",
    doi = "10.1007/JHEP01(2019)026",
    journal = "JHEP",
    volume = "01",
    pages = "026",
    year = "2019"
}

@article{Petkova:2000ip,
    author = "Petkova, V. B. and Zuber, J. B.",
    title = "{Generalized twisted partition functions}",
    eprint = "hep-th/0011021",
    archivePrefix = "arXiv",
    doi = "10.1016/S0370-2693(01)00276-3",
    journal = "Phys. Lett. B",
    volume = "504",
    pages = "157--164",
    year = "2001"
}

@article{Frohlich:2004ef,
    author = {Fr{\"o}hlich, J{\"u}rg and Fuchs, J{\"u}rgen and Runkel, Ingo and Schweigert, Christoph},
    title = "{Kramers-Wannier duality from conformal defects}",
    eprint = "cond-mat/0404051",
    archivePrefix = "arXiv",
    doi = "10.1103/PhysRevLett.93.070601",
    journal = "Phys. Rev. Lett.",
    volume = "93",
    pages = "070601",
    year = "2004"
}

@article{Loladze:2024ayk,
    author = "Loladze, Vazha and Okui, Takemichi",
    title = "{Monopole-Fermion Scattering and the Solution to the Semiton/Unitarity Puzzle}",
    eprint = "2408.04577",
    archivePrefix = "arXiv",
    primaryClass = "hep-th",
    doi = "10.1103/PhysRevLett.134.051602",
    journal = "Phys. Rev. Lett.",
    volume = "134",
    number = "5",
    pages = "051602",
    year = "2025"
}

@article{Ohmori:2026tip,
    author = "Ohmori, Kantaro and Watanabe, Masataka and Wei, Pengcheng and Zheng, Yunqin",
    title = "{Intertwining Defects and Fermion-Boundary Scattering from TQFT}",
    eprint = "2610.31737",
    archivePrefix = "arXiv",
    primaryClass = "hep-th",
    reportNumber = "RIKEN-iTHEMS-Report-26",
    month = "10",
    year = "2026"
}

@article{Koundouros:2026rnb,
    author = "Koundouros, George and Tong, David",
    title = "{Fermions Scattering Off Non-Abelian Rotors}",
    eprint = "2610.31780",
    archivePrefix = "arXiv",
    primaryClass = "hep-th",
    month = "10",
    year = "2026"
}

@article{Arias-Tamargo:2026kbv,
    author = "Arias-Tamargo, Guillermo and Boyle Smith, Philip and Moulan, Rishi and Vel{\'a}squez Cotini Hutt, Max",
    title = "{Twisted scattering from Chern-Simons theory}",
    eprint = "2610.30232",
    archivePrefix = "arXiv",
    primaryClass = "hep-th",
    reportNumber = "IFT-UAM/CSIC-26-155, Imperial-TP-2026-RM-1",
    month = "10",
    year = "2026"
}

@article{Lin:2023uvm,
  author = {Lin, Ying-Hsuan and Shao, Shu-Heng},
  title = {{Bootstrapping noninvertible symmetries}},
  eprint = {2302.13900},
  archiveprefix = {arXiv},
  primaryclass = {hep-th},
  reportnumber = {YITP-SB-2023-03},
  doi = {10.1103/PhysRevD.107.125025},
  journal = {Phys. Rev. D},
  volume = {107},
  number = {12},
  pages = {125025},
  year = {2023},
}

@article{Lin:2021udi,
    author = "Lin, Ying-Hsuan and Shao, Shu-Heng",
    title = "{$\mathbb{Z}_N$ symmetries, anomalies, and the modular bootstrap}",
    eprint = "2101.08343",
    archivePrefix = "arXiv",
    primaryClass = "hep-th",
    reportNumber = "CALT-TH-2021-002",
    doi = "10.1103/PhysRevD.103.125001",
    journal = "Phys. Rev. D",
    volume = "103",
    number = "12",
    pages = "125001",
    year = "2021"
}

@article{Lin:2019kpn,
    author = "Lin, Ying-Hsuan and Shao, Shu-Heng",
    title = "{Anomalies and Bounds on Charged Operators}",
    eprint = "1904.04833",
    archivePrefix = "arXiv",
    primaryClass = "hep-th",
    reportNumber = "CALT-TH-2019-009",
    doi = "10.1103/PhysRevD.100.025013",
    journal = "Phys. Rev. D",
    volume = "100",
    number = "2",
    pages = "025013",
    year = "2019"
}

@article{Thorngren:2021yso,
    author = "Thorngren, Ryan and Wang, Yifan",
    title = "{Fusion category symmetry. Part II. Categoriosities at c = 1 and beyond}",
    eprint = "2106.12577",
    archivePrefix = "arXiv",
    primaryClass = "hep-th",
    doi = "10.1007/JHEP07(2024)051",
    journal = "JHEP",
    volume = "07",
    pages = "051",
    year = "2024"
}

@article{Bhardwaj:2017xup,
  author = {Bhardwaj, Lakshya and Tachikawa, Yuji},
  title = {{On finite symmetries and their gauging in two dimensions}},
  eprint = {1704.02330},
  archiveprefix = {arXiv},
  primaryclass = {hep-th},
  reportnumber = {IPMU-17-0049},
  doi = {10.1007/JHEP03(2018)189},
  journal = {JHEP},
  volume = {03},
  pages = {189},
  year = {2018},
}

@article{Callias:1977cc,
    author = "Callias, Constantine J.",
    title = "{Spectra of Fermions in Monopole Fields: Exactly Soluble Models}",
    reportNumber = "MIT-CTP-624",
    doi = "10.1103/PhysRevD.16.3068",
    journal = "Phys. Rev. D",
    volume = "16",
    pages = "3068",
    year = "1977"
}

@article{Goldhaber:1977xw,
    author = "Goldhaber, Alfred S.",
    title = "{Dirac Particle in a Magnetic Field: Symmetries and their Breaking by Monopole Singularities}",
    reportNumber = "ITP-SB-77-29",
    doi = "10.1103/PhysRevD.16.1815",
    journal = "Phys. Rev. D",
    volume = "16",
    pages = "1815",
    year = "1977"
}

@article{Csaki:2020inw,
    author = "Csaki, Csaba and Hong, Sungwoo and Shirman, Yuri and Telem, Ofri and Terning, John and Waterbury, Michael",
    title = "{Scattering amplitudes for monopoles: pairwise little group and pairwise helicity}",
    eprint = "2009.14213",
    archivePrefix = "arXiv",
    primaryClass = "hep-th",
    doi = "10.1007/JHEP08(2021)029",
    journal = "JHEP",
    volume = "08",
    pages = "029",
    year = "2021"
}

@article{Csaki:2024ajo,
    author = "Cs{\'a}ki, Csaba and Ovadia, Rotem and Telem, Ofri and Terning, John and Yankielowicz, Shimon",
    title = "{Abelian instantons and monopole scattering}",
    eprint = "2406.13738",
    archivePrefix = "arXiv",
    primaryClass = "hep-th",
    doi = "10.1007/JHEP11(2024)165",
    journal = "JHEP",
    volume = "11",
    pages = "165",
    year = "2024"
}

@article{Csaki:2021ozp,
    author = "Cs{\'a}ki, Csaba and Shirman, Yuri and Telem, Ofri and Terning, John",
    title = "{Pairwise Multiparticle States and the Monopole Unitarity Puzzle}",
    eprint = "2109.01145",
    archivePrefix = "arXiv",
    primaryClass = "hep-th",
    doi = "10.1103/PhysRevLett.129.181601",
    journal = "Phys. Rev. Lett.",
    volume = "129",
    number = "18",
    pages = "181601",
    year = "2022"
}

@article{Kitano:2021pwt,
    author = "Kitano, Ryuichiro and Matsudo, Ryutaro",
    title = "{Missing final state puzzle in the monopole-fermion scattering}",
    eprint = "2103.13639",
    archivePrefix = "arXiv",
    primaryClass = "hep-th",
    reportNumber = "KEK-TH-2313, KEK-TH-2321",
    doi = "10.1016/j.physletb.2022.137271",
    journal = "Phys. Lett. B",
    volume = "832",
    pages = "137271",
    year = "2022"
}

\end{document}